\documentclass[12pt,journal,onecolumn,draftcls]{IEEEtran}

\usepackage[utf8]{inputenc} 
\usepackage[T1]{fontenc}
\usepackage{url}
\usepackage{ifthen}
\usepackage{cite}
\usepackage[cmex10]{amsmath} 

\usepackage{times}

\usepackage{amstext,amssymb,bm}
\usepackage{latexsym}
\usepackage{pstricks}

\usepackage{enumerate}
\usepackage{mathtools}

\usepackage{epsfig}
\usepackage{times}
\usepackage{float}
\usepackage{afterpage}
\usepackage{amsmath}
\usepackage{amstext}
\usepackage{amssymb}
\usepackage{amssymb,bm}
\usepackage{latexsym}
\usepackage{color}
\usepackage{amsmath}
\usepackage{amsthm}
\usepackage{graphicx}
\usepackage{pstricks}
\usepackage{booktabs}
\usepackage{enumerate}

\usepackage{fixltx2e}
\usepackage{enumitem}

\usepackage{multirow}

\usepackage[ruled,vlined]{algorithm2e}

\DeclarePairedDelimiter\ceil{\lceil}{\rceil}
\DeclarePairedDelimiter\floor{\lfloor}{\rfloor}

\newtheorem{thm}{Theorem}
\newtheorem*{thm*}{Theorem}
\newtheorem{lemma}{Lemma}

\newtheorem{prop}{Proposition}
\newtheorem{rem}{Remark}

\newtheorem{example}{Example}

\usepackage{verbatim}
\usepackage{mathabx}
\usepackage{hyperref}
\usepackage{subcaption}

\newcommand{\mais}{\text{MAIS}}
\newcommand{\smin}{s_{\rm min}}
\newcommand{\smax}{s_{\rm max}}
\newcommand{\so}{s_{\rm min}}
\newcommand{\st}{s_{\rm max}}
\newcommand{\nth}{network topology hypergraph}
\newcommand{\cah}{circular-arc hypergraph}
\newcommand{\canth}{circular-arc network topology hypergraph}

\newcommand{\cardi}{t}
\newcommand{\mbit}{\kappa}

\begin{document}
\title{Tight Information Theoretic Converse Results for some Pliable Index Coding Problems} 

\author{%
\IEEEauthorblockN{%
Tang Liu and Daniela Tuninetti\\%
University of Illinois at Chicago, Chicago, IL 60607 USA, 
Email: {\tt tliu44, danielat@uic.edu}\\%
}%
}
\maketitle


\footnotetext{%
This work was presented in part at ITW 2017 and ITW 2018.
The work of the authors was partially funded by NSF under award number 1527059.
The contents of this article are solely the responsibility of the author and do not necessarily represent the official views of the NSF.}

\begin{abstract}
This paper studies the Pliable Index CODing problem (PICOD), which models content-type distribution networks. 
In the PICOD$(\cardi)$ problem there are $m$ messages, $n$ users and each user has a distinct message side information set, as in the classical Index Coding problem (IC). Differently from IC, where each user has a pre-specified set of messages to decode, in the PICOD$(\cardi)$ a user is ``pliable'' and is satisfied if it can decode any $\cardi$ messages that are not in its side information set. The goal is to find a code with the shortest length that satisfies all the users. This flexibility in determining the desired message sets makes the PICOD$(\cardi)$ behave quite differently compared to the IC, and its analysis challenging.

This paper mainly focuses on the \emph{complete--$S$} PICOD$(\cardi)$ with $m$ messages, where the set $S\subset[m]$ contains the sizes of the side information sets, and the number of users is $n=\sum_{s\in S}\binom{m}{s}$, with no two users having the same side information set. 
Capacity results are shown for:
(i)  the \emph{consecutive} complete--$S$ PICOD$(\cardi)$, where $S=[\smin:\smax]$ for some $0 \leq \smin\leq \smax \leq m-\cardi$, and
(ii) the \emph{complement-consecutive} complete--$S$ PICOD$(\cardi)$, where $S=[0:m-\cardi]\backslash[\smin:\smax]$,  
for some $0 < \smin\leq \smax < m-\cardi$.
The novel converse proof is inspired by combinatorial design techniques and the key insight is to consider all messages that a user can eventually decode successfully, even those in excess of the $\cardi$ required ones.
This allows one to circumvent the need to consider all possible desired message set assignments at the users in order to find the one that leads to the shortest code length. 
The core of the novel proof is to solve the \emph{critical} complete--$S$ PICOD$(\cardi)$ with $m = 2s+\cardi$ messages and $S=\{s\}$, by showing the existence of a user who can decode $s+\cardi$ messages regardless of the desired message set assignment. 
All other tight converse results for the complete--$S$ PICOD$(\cardi)$ can be deduced from this critical case.
The converse results show the information theoretic optimality of simple linear coding schemes.
By similar reasoning, all complete--$S$ PICOD$(\cardi)$ where the number of messages is $m\leq 5$ can be solved.

In addition, tight converse results are also shown for those PICOD$(1)$ with \canth.

\end{abstract}

\section{Introduction} 
\label{sec:introduction}

\subsection{Motivation}
The broadcast channel with message side information at the receivers has became a critical model to understand the full potential of wireless communication networks as it models, for example, the downlink of the two-way relay channel~\cite{two_wayrelay}. Not even the capacity of the general broadcast channel without receiver side information is known. Therefore, some practically motivated and reasonably simple models are of interest when message side information at the receivers is considered.
Index coding (IC) is one such model. First proposed in~\cite{index_coding_original} when considering satellite communication, the IC consists of one transmitter with $m$ independent messages to be delivered to $n$ users through an error-free broadcast link. Each user has some messages as side information available to it and needs to reliably decode some messages that are not in its side information set; the desired messages for each user are pre-determined. In IC, one asks what is the minimum number of transmissions (i.e., minimum code length) such that every user is able to decode its desired messages successfully~\cite{index_coding_with_sideinfo}. 
Compared to the general broadcast channel with side information at the users, the IC appears simple because: 
1) the channel is noiseless, and
2) the side information sets are proper subsets of the whole message set.
The IC focuses on the benefits / transmitter encoding opportunities brought by the different side information sets at the users.
However, the general IC is still open.
When one restricts attention to linear codes, the optimal code length is fully characterized by the so-called \emph{minrank} problem, which is NP-complete in general~\cite{index_coding_with_sideinfo}.
In~\cite{IC_NC_matroid} it is proved that the IC, which is a special network coding problem, is in fact equivalent to the general network coding problem. 
Therefore, as for network coding, for IC linear schemes are not sufficient~\cite{linear_suboptimality_IC} and non-Shannon type of inequality are necessary~\cite{IC_nonshannon}.

The IC problem models scenarios where the transmitter can do encoding based on the side information sets and on fixed desired message sets for the users.
In practice, there may be flexibility in choosing the desired message sets. 
For example, in a music streaming service, users do not know which song will be played next; they are usually only guaranteed that it will be one from a certain group and not repeated.
In online advertisement systems, the clients do not require a specific advertisement to see; it is the distributor who chooses what will be put on the clients' screens; the distributor might want to avoid repeating the same advertisement at the same client, as it might decrease the client's satisfaction.
These scenarios can be modeled as a variant of the IC where the users are satisfied by \emph{any} message that is not in its side information set, instead of a specific one as in the original IC setting. The transmitter thus has the freedom to choose the messages conveyed to the users so to minimize the transmission duration, or code length.

In this paper, we study this variant of IC known as Pliable Index CODing (PICOD), firstly proposed in~\cite{BrahmaFragouli-IT1115-7254174}.
The PICOD$(\cardi)$ and the IC share many attributes.
In the PICOD$(\cardi)$, one still has a single transmitter with $m$ message and $n$ users with message side information.  
The transmitter and users are connected via a shared noiseless rate-limited broadcast channel.
The only major difference is that for the PICOD$(\cardi)$ the desired message sets at the users are not pre-determined and each user is satisfied whenever it can decode \emph{any} $\cardi$ messages not in its side information set.
This provides the transmitter more encoding opportunities, as it now encodes based on its own choice of desired message sets for the users, by knowing the message side information sets at each user.
The goal in the PICOD$(\cardi)$ is to find the desired message set assignment that leads to the smallest possible code length.

\subsection{Past Work on PICOD} 
As one would expect, the extra freedom of choosing the desired message sets in the PICOD$(\cardi)$ significantly reduces the number of transmissions / code length compared to the classical IC with the same number of messages, number of users, and message side information sets.
In~\cite{BrahmaFragouli-IT1115-7254174}, when all side information sets are of size $s\leq m-\cardi$, it showed that there exits a code of length $O(\min\{t\log (n), t+\log^2 (n)\})$ for the PICOD$(\cardi)$.
When there is no constraint on the size of side information sets, and $m=O(n^\delta)$ for some constant positive $\delta$, a code length of $O(\min\{t\log^2 (n),t\log (n)+\log^3 (n)\})$ is achievable~\cite{BrahmaFragouli-IT1115-7254174}.
Recently in~\cite{polytime_alg_picod}, a deterministic polynomial time algorithm was proposed to achieve a code length of $O(\log^2 (n))$ for $t=1$ and of $O(t\log(n)+\log^2(n))$ otherwise.
Those results show an exponential code length reduction for the PICOD$(t)$ compared to the IC~\cite{BrahmaFragouli-IT1115-7254174}.

An interesting model proposed in~\cite{BrahmaFragouli-IT1115-7254174} is the so-called \emph{oblivious} PICOD$(\cardi)$.
In the oblivious PICOD$(\cardi)$ the transmitter does not know the specific side information sets at the users.
The transmitter only has knowledge of the sizes of the side information sets.
In~\cite{BrahmaFragouli-IT1115-7254174,constant_frac_satisfactory} the authors proved that for the oblivious PICOD$(\cardi)$ at least a fraction $1/e$ of the remained unsatisfied users can be satisfied at each new transmission. 
This shows that there exists an achievable scheme where the code length is the logarithm of the number of users in the system, 
which is an exponential improvement in the number of transmissions compared to the IC.

Known achievable schemes for the PICOD$(\cardi)$ are based on linear codes only, and very few converse results are available. 
To the best of our knowledge, all converse proofs show 
bounds under the constraint that the code used is linear. 
For the oblivious PICOD$(\cardi)$, the optimal code length under the restriction that the transmitter can only use linear schemes is shown in~\cite[Theorem~9]{BrahmaFragouli-IT1115-7254174}.
In~\cite{polytime_alg_picod}, the authors provide a worst case instance that needs $\Omega(\log (n))$ code length for linear codes. The objective of this paper is to prove information theoretic converse results for some classes of the PICOD$(\cardi)$ without any restrictions of the class of codes used at the transmitter.


\subsection{Contributions}
In this paper we derive tight information theoretic converse bounds for some PICOD$(\cardi)$ problems based on the structure of the side information sets, namely:
(i)  the \emph{complete-$S$} PICOD$(\cardi)$, and
(ii) the PICOD$(\cardi)$ with a \emph{\canth}.


The complete--$S$ PICOD$(\cardi)$, where $S$ is a subset of $[0:m-\cardi]$ (where $m$ is the number of messages at the transmitter and $t$ the number of messages to be decoded), is a system where all side information sets / users with size indexed by $S$ are present.
We say that $S$ is \emph{consecutive} if $S=[\smin: \smax]$ for some $0\leq \smin\leq \smax\leq m-\cardi$, which is also known as the oblivious PICOD$(\cardi)$ in~\cite{BrahmaFragouli-IT1115-7254174}.
In~\cite{BrahmaFragouli-IT1115-7254174} the authors derived tight converse bounds for the oblivious PICOD$(\cardi)$ when the coding scheme is restricted to be a linear code.
In this work, we aim to provide tight information theoretic converse bounds, i.e., without any restriction on the coding scheme being used, on the same model. Our complete--$S$ PICOD$(\cardi)$ setting actually includes and expands on the oblivious PICOD$(\cardi)$ setting studied in~\cite{BrahmaFragouli-IT1115-7254174}, and our results show the unrestricted optimality of linear codes.

Our converse is based on showing the existence of at least one \emph{special user} who can decode a certain number of messages outside its side information set; the stumbling block in previous approaches was how to find such a special user.
The problem of finding the special user can be approached in two ways: 
1) constructively finding such a special user for each choices of desired messages, or 
2) implicitly proving its existence.
In this work we use both methods.

\paragraph*{Constructive Method} 
For the \emph{complement-consecutive} complete--$S$ PICOD$(\cardi)$, which is the complete--$S$ PICOD$(\cardi)$ with $S=[0:m-\cardi]\setminus [\so:\st]$ where $0 < \so \leq \st < m-\cardi$, we constructively find the special user that can decode $|S|+t-1$ messages, i.e., the one whose side information set is empty.

\paragraph*{Combinatorial Method} 
The constructive method is not amenable for the consecutive complete--$S$ PICOD$(\cardi)$, which is the complete--$S$ PICOD$(\cardi)$ with $S=[\smin : \smax]$ where $0 \leq \smin\leq \smax \leq m-\cardi$, due to the large number of sub-cases / different desired message set assignments that must be considered separately.
Therefore for this case we propose a novel combinatorial proof to show the existence of a special user.
By not only focusing on the desired messages, but on all the messages that a user is eventually able to decode, we consider the messages that a user can eventually know as a \emph{block cover} for this user's side information set;
the terminology is borrowed fom the combinatorial design structure known as Steiner system~\cite{generalized_steiner_system}.
We argue that the absence of a special user leads to a contradiction in this block cover, and that therefore a special user must exist. 
This new technique greatly reduces the complexity of the proof compared to the constructive method and enables us to obtain a converse bound for a very general class of complete--$S$ PICOD$(\cardi)$ problems. The keystone of the proof is to show that, for the \emph{critical} complete--$S$ PICOD$(\cardi)$ case with $S=\{s\}$ and $m = 2s+\cardi$, there must exist at least one user who can decode $s+\cardi$ messages. 
From this, the extension to the consecutive complete--$S$ PICOD$(\cardi)$ follows by enhancing the system to a critical one.
By similar reasoning, all complete--$S$ PICOD$(\cardi)$ where the number of messages is $m\leq 5$ can be solved.

The idea of showing the existence of a special user can also be used for the following PICOD$(\cardi)$ problem--for a detailed definition please refers to Section~\ref{sub:graph_prelimiary}.
For the case $\cardi=1$ we show a tight converse for those PICOD$(1)$ with \canth. For this setting, when a $1$-factor does not exist we show that the code length is at least two by finding a user that can decode two messages. 

\subsection{Paper Organization}
The rest of the paper is organized as follows: 
Section~\ref{sec:system_model} introduces the system model and related definitions;
Section~\ref{sec:main_results} presents the main results of this paper;
Sections~\ref{sec:layer_counting_converse}-\ref{sec:complete_nonconsecutive} present converse proofs for some complete--$S$ PICOD$(\cardi)$ problems and their optimality; 
Section~\ref{sec:picod_with_special_network_topology_hypergraph} shows the optimal information theoretic converse for the PICOD$(1)$ with \canth;
Section~\ref{sec:conclusion} concludes the paper and discusses future work;
some proofs can be found in Appendix.

\subsection{Notation}
\label{sub:notation}
Throughout the paper we use 
capital letters to denote sets,
calligraphic letters for family of sets, and 
lower case letters for elements in a set. 
The cardinality of the set $A$ is denoted by $|A|$. 
For integers $a_1, a_2$ 
we let $[a_1:a_2] := \{a_1,a_1+1,\ldots,a_2\}$ for $a_1\leq a_2$
and $[a_2]:=[1:a_2]$ for $a_2\geq 1$.
A capital letter as a subscript denotes set of elements whose indices are in the set, i.e., $W_A:=\{w_a : w\in W, a\in A\}$.
For two sets $A$ and $B$, $A\setminus B$ is the set that consists all the elements that are in $A$ but not in $B$.
Notations and nomenclature from graph theory will be introduced in Section~\ref{sec:picod_with_special_network_topology_hypergraph}.

\section{System Model} 
\label{sec:system_model}

In a PICOD$(\cardi)$ system there is one server / transmitter and $n\in\mathbb{N}$ clients / users; 
the user set is denoted as $U := \left\{ u_{1},u_{2},\ldots,u_{n}\right\}$. 
The server is connected to all users via a rate-limited noiseless broadcast channel. 
There are $m\in\mathbb{N}$ independent and uniformly distributed binary messages of $\mbit \in \mathbb{N}$ bits each; 
the message set is denoted as $W := \left\{ w_{1},w_{2},\ldots,w_{m} \right\}$.
User $u_i$ has a subset of the message set as its side information set $A_i\subset [m]$, $i\in[n]$. 
The collection of all side information sets is denoted as $\mathcal{A} := \{A_{1},A_{2},\ldots,A_{n}\}$;
$\mathcal{A}$ is assumed globally known at the transmitter and all users. 

The server broadcasts to the users a codeword of length $\ell \mbit$ bits, which is a function of the message set $W$ and the collection of all side information sets $\mathcal{A}$, i.e., for some function $\mathsf{ENC}$ we have
\begin{align}
	x^{\ell \mbit} = \mathsf{ENC}(W,\mathcal{A}).	
\end{align}
Each user decodes based on the received $x^{\ell \mbit}$ and its own side information set;
for user $u_j, j\in[n]$, the decoding function is 
\begin{align}
	\{\widehat{w}^{(j)}_{1},\dots,\widehat{w}^{(j)}_{\cardi}\} = \mathsf{DEC}_j(W_{A_j},x^{\ell \mbit}).
\end{align}
A code is said to be \emph{valid} if and only if every user can successfully decode at least $\cardi$ messages not in its side information set, i.e., the decoding functions $\{ \mathsf{DEC}_j, \forall j\in[n]\}$ are such that
\begin{align}
	\Pr\left[ \exists \{d_{j,1},\dots,d_{j,\cardi}\}\cap A_j=\emptyset : \{\widehat{w}^{(j)}_{1},\dots,\widehat{w}^{(j)}_{\cardi}\}\neq \{{w}_{d_{j,1}},\dots,{w}_{d_{j,\cardi}}\} \ \text{for some $j\in[n]$} \right]  \leq \epsilon,
\end{align}
for some $\epsilon\in(0,1)$.
For a valid code,  
$\{\widehat{w}^{(j)}_{1},\dots,\widehat{w}^{(j)}_{\cardi}\} = \{{w}_{d_{j,1}},\dots,{w}_{d_{j,\cardi}}\}$ is called the \emph{desired message set} for user $u_j, \ j\in[n],$ and the indices of the desired messages are denoted as $D_j:=\{d_{j,1},\dots,d_{j,\cardi}\}$  where $D_j\cap A_j=\emptyset, \forall j\in [n]$.
The choice of desired messages for the users is denoted as $\mathcal{D}=\{D_1,D_2,\ldots\,D_n\}$.
The goal is to find a valid code with minimum length
\begin{align}
	\ell^*:= \min\{\ell : \text{$\exists$ a valid code of length $\ell \mbit$, for some $\mbit$}\}.	
\end{align}


In the following we shall mainly focus on the \emph{complete--$S$ PICOD$(\cardi)$}, for a given set $S\subseteq[0:m-\cardi]$.
In this system, there are $n := \sum_{s\in S}\binom{m}{s}$ users, where no two users have the same side information set. 
In other words, all possible users with distinct side information sets that are subsets of size $s$ of the message set, for all $s\in S$, are present in the complete--$S$ PICOD$(\cardi)$.

\section{Main Results and Discussion} 
\label{sec:main_results}
This section summarizes our main results and comments on their proof techniques, their relationship with past work, and their implications.
We start with a simple achievable scheme based on linear codes, in Section~\ref{subsec:ach}. 
The main contribution of the paper is converse bounds on the optimal code length for the two broad families of PICOD$(\cardi)$: 
(i)  some complete-$S$ PICOD$(\cardi)$, where $S$ is nonempty subset of $[0:m-\cardi]$, in Section~\ref{subsec:converse-complete}, and
(ii) the PICOD$(1)$ with \canth\ in Section~\ref{subsec:converse-hyper}.

\subsection{Achievability}
\label{subsec:ach}
We give next an achievable scheme for the general complete--$S$ PICOD$(\cardi)$ based on linear codes.
\begin{prop}[Achievable Scheme]
	\label{prop:achievability_complete-s}
	Let $\mathcal{S}$ by a partition of $S$, i.e., $S=\cup_{i\in [|\mathcal{S}|]}S_i$ and 
	$S_i\cap S_j =\emptyset$ for all $i,j\in [|\mathcal{S}|]$ such that $i\neq j$.
	The optimal code length for the complete--$S$ PICOD$(\cardi)$ with $m$ messages is upper bounded by
	\begin{align}
		\ell^*\leq  \sum_{i\in [|\mathcal{S}|]}\min\left\{m-\min_{s\in S_i}\{s\}, \max_{s\in S_i}\{s\}+\cardi\right\}.
		\label{eq:prop:achievability_complete-s qaz}
	\end{align}
	By minimizing over all possible partitions $\mathcal{S}$, 
	we have
	\begin{align}
		\ell^*\leq \min_{\mathcal{S}} \sum_{i\in [|\mathcal{S}|]}\min\left\{m-\min_{s\in S_i}\{s\}, \max_{s\in S_i}\{s\}+\cardi\right\}.
		\label{eq:prop:achievability_complete-s}
	\end{align}
\end{prop}
The proof is simple and can be deduced from Remark~\ref{rem:achievability}. 
\begin{rem}
	\label{rem:achievability}
	Proposition~\ref{prop:achievability_complete-s} is a generalization of the 
	scheme proposed in~\cite{BrahmaFragouli-IT1115-7254174} whose main idea is as follows.
	Let $\smin$ and $\smax$ denote the smallest and largest 
	size of the side information sets, respectively.
	Transmitting $\smax+\cardi$ messages one by one can satisfy all users since each user has at most $\smax$ messages in its side information set.
	Transmitting $m-\smin$ linearly independent linear combinations of the $m$ messages also satisfies all users, as each user has at least $\smin$ messages in its side information set. 
	Therefore by choosing the best of these two linear codes, 
	we have $\ell^*\leq \min\{\smax+\cardi, m-\smin\}$.
	
	We generalize this idea for the complete--$S$ PICOD$(\cardi)$ by partitioning $S$ into the collection $\mathcal{S}$ and by satisfying the users in each $S_i\in\mathcal{S}$ by using the above scheme.
	The total code length is the sum of the length of the code used in each partition.
	Finally, the shortest code length this scheme can achieve is given by searching the best possible partition of $S$.
\end{rem}

\subsection{Converse for some complete--$S$ PICOD$(\cardi)$ problems}
\label{subsec:converse-complete}
We show that for two choices of $S$ the achievability in Proposition~\ref{prop:achievability_complete-s} is information theoretic optimal. 

\begin{thm}[Converse for the \emph{complement-consecutive} complete--$S$ PICOD$(\cardi)$]
	\label{thm:layer_cont_opt}
	For the complete--$S$ PICOD$(\cardi)$ with $m$ messages and $S=[0:m-\cardi]\backslash [\smin:\smax] = [0:\so-1]\cup[\st+1:m-\cardi]$ for some $0 < \so\leq \st < m-\cardi$ (note that the set $S$ includes elements $0$ and $m-\cardi$), the optimal code length is
	\begin{align}
	\ell^*=\min\{m,m+\cardi+\so-\st-2\}=\min\{m,|S|+2\cardi-2\}.
	\label{eq:thm:layer_cont_opt}
	\end{align}
\end{thm}
The proof of Theorem~\ref{thm:layer_cont_opt} can be found in Section~\ref{sec:layer_counting_converse}.

\begin{thm}[Converse for the \emph{consecutive} complete--$S$ PICOD$(\cardi)$]
	\label{thm:complete_consecutive}
	For the complete--$S$ PICOD$(\cardi)$ with $m$ messages and $S=[\smin:\smax]$ for some $0\leq \smin \leq \smax \leq m-\cardi$ (i.e., $S$ contains consecutive integers, from $\smin$ to $\smax$) the optimal code length is
	\begin{align}
	\ell^*=\min\{\smax+\cardi, m-\smin\}.
	\label{eq:thm:complete_consecutive}
	\end{align}
\end{thm}
The proof of Theorem~\ref{thm:complete_consecutive} is broken down in several pieces. The proof for the \emph{critical case}, where $m=2s+\cardi$ and $S=\{s\}$, can be found in Section~\ref{sec:complete_s_picod_m=2s+1}, while the general proof in Section~\ref{sec:complete_s_consecutive}.

\begin{rem}
 \label{rem:converse}
	Theorems~\ref{thm:layer_cont_opt} and~\ref{thm:complete_consecutive} show that the simple achievable scheme in Proposition~\ref{prop:achievability_complete-s} can be information theoretical optimal for a class of PICOD$(\cardi)$. Specifically, the consecutive complete--$S$ PICOD$(\cardi)$ is the oblivious PICOD$(\cardi)$ studied in~\cite{BrahmaFragouli-IT1115-7254174}. Our Theorem~\ref{thm:complete_consecutive} provides a tight information theoretic converse for the achievability proposed in~\cite{BrahmaFragouli-IT1115-7254174}.
	
	The basic idea in the proof of Theorem~\ref{thm:layer_cont_opt} is 
	to prove the existence of a user who can decode $|S|$ messages by
	a method referred to as \emph{layer counting}. 
	We partition all users in the complete--$S$ PICOD$(\cardi)$ into $|S|$ layers.
	Each layer contains the users with the same size of the side information set. 
	A layer is said to be ``lower'' than another if the size of the side information set of the users is smaller.
	The intuition is that a user in a lower layer, after having decoded its desired messages,  
	can mimic users in higher layers and thus decode also the desired messages of those higher layer users. 

	In the complement-consecutive complete--$S$ PICOD$(\cardi)$, where $S=[0:\so-1]\cup[\st+1:m-\cardi]$ for some $0< \so\leq \st < m-\cardi$, we show the user in the lowest layer (with empty side information set) can mimic a user in each higher layers and eventually decodes $|S|+2\cardi-2$ messages. 

	However, this layer counting converse is not tight in general, as explained in Remark~\ref{rem:layer counting not tight}  
	for the complete--$S$ PICOD$(1)$ with $S=[1:q]$ or $S=[q:m-2]$ for some $2\leq q \leq m-2$.
	To improve on the layer counting converse, we propose a novel converse technique in Theorem~\ref{thm:complete_consecutive} for the \emph{consecutive} complete--$S$ PICOD$(\cardi)$, where $S=[\smin: \smax]$ for some $0\leq \smin \leq \smax \leq m-\cardi$. 
	The \emph{critical case} for this proof is the complete--$S$ PICOD$(\cardi)$ for 
	\begin{align}
	\text{$m=2s+\cardi$ messages and $S=\{s\}$ \ \ (critical case)}. 
	\label{eq:critical}
	\end{align}
	In Section~\ref{sec:complete_s_picod_m=2s+1} Proposition~\ref{prop:power_user_exist}, we show that for this critical case, regardless of the choice of desired messages and valid code, there always exists at least one user who can decode $s+\cardi$ messages.
	While the proof of Theorem~\ref{thm:layer_cont_opt} is constructive, that is, we explicitly identify the user who can always decode $|S|+2\cardi-2$ messages (the one with empty side information set), the proof of Proposition~\ref{prop:power_user_exist} is not.
	The problem with a constructive argument for the critical case is that, 
	for any specific user, there exists an information theoretic optimal choice of desired messages and a corresponding valid code such that this user can decode only its desired $\cardi$ messages and no more.
	In other words, showing that a certain user can always decode more than $\cardi$ messages is impossible.
	Therefore, in the proof of Proposition~\ref{prop:power_user_exist}, we propose a combinatorial method to show the existence of at a least a user with some desired property, namely, the ability to decode a certain number of messages.
	The new method involves the Maximum Acyclic Induced Subgraph (MAIS) converse idea for the classic IC~\cite{index_coding_with_sideinfo}, as well as a combinatorial design technique inspired by Steiner systems~\cite{generalized_steiner_system},  which we shall refer to as \emph{block cover}.
	The existence proof does not indicate which user has the desired property, but only shows its existence regardless of the choice of desired messages at the users.
\end{rem}


Theorem~\ref{thm:complete_consecutive} can be further extended to cover other complete--$S$ PICOD$(\cardi)$.  
We have the following results.

\begin{prop}[Not a complete--$S$ system, but all users are below the critical case users in the layer representation]
\label{prop:max<m/2_nonconsecutive}
	For the complete--$S$ PICOD$(\cardi)$ with $m$ messages and $\smax :=\max_{s\in S}\{s\} \leq \floor{\frac{m-\cardi}{2}}$, the optimal code length is $\ell^*=\smax+\cardi$.
\end{prop}
The proof can be found in Section~\ref{sec:complete_nonconsecutive}.

\begin{prop}[Not a complete--$S$ system, but all users are above the critical case users in the layer representation]
\label{prop:min>m/2_nonconsecutive}
	For the complete--$S$ PICOD$(\cardi)$ with $m$ messages and $\smin :=\min_{s\in S}\{s\}\geq \ceil{\frac{m-\cardi}{2}}$, the optimal code length is $\ell^*=m-\smin$.
\end{prop}
The proofs can be found in Section~\ref{sec:complete_nonconsecutive}.

\begin{prop}[Not a complete--$S$ system, but all users in a band around the critical case users are present in the layer representation]
\label{prop:min<m/2<max_nonconsecutive}
	For the complete--$S$ PICOD$(\cardi)$ with $m$ messages, let
	\begin{align}
	\delta :=\min\left\{\smax-\ceil{\frac{m-\cardi}{2}}, \floor{\frac{m-\cardi}{2}}-\smin\right\}, 
	\label{eq:def delta band}
	\end{align}
	where $\smax:=\max_{s\in S}\{s\}$ and $\smin:=\min_{s\in S}\{s\}$. 
	If $\left[\floor{\frac{m-\cardi}{2}}-\delta:\ceil{\frac{m-\cardi}{2}}+\delta\right]\subseteq S$ then the optimal code length is $\ell^*=\min\{\smax+\cardi,m-\smin\}$.
\end{prop}
The proof can be found in Section~\ref{sec:complete_nonconsecutive}.

\begin{rem}
	Propositions~\ref{prop:max<m/2_nonconsecutive},~\ref{prop:min>m/2_nonconsecutive} and~\ref{prop:min<m/2<max_nonconsecutive} show an interesting fact: for these settings the only relevant layers in the layer representation are the ones closest to the ``critical'' middle layer $\frac{m-\cardi}{2}$, or the layers in a band $\left[\floor{\frac{m-\cardi}{2}}-\delta:\ceil{\frac{m-\cardi}{2}}+\delta\right]$ around the ``critical'' middle layer. The optimal code for the users in these layers satisfies all the remaining users. 
\end{rem}

Finally, for those PICOD$(\cardi)$ problems with $m\leq 5$ messages that are not covered by Propositions~\ref{prop:max<m/2_nonconsecutive},~\ref{prop:min>m/2_nonconsecutive},~\ref{prop:min<m/2<max_nonconsecutive} and Theorem~\ref{thm:layer_cont_opt},  we have the following:
\begin{prop}
	\label{prop:achi_opt_m<=5}
	For all complete--$S$ PICOD$(\cardi)$ with $m\leq 5$ and non-empty $S\subseteq[0:m-1]$, 
	the achievable scheme in Proposition~\ref{prop:achievability_complete-s} is information theoretic optimal.
\end{prop}
The proof can be found in Section~\ref{sec:complete_nonconsecutive}.

\begin{rem}
	Proposition~\ref{prop:achi_opt_m<=5} is proved by checking one by one all complete--$S$ PICOD$(\cardi)$ problems with $m\leq 5$ messages not covered by previous results. It may be possible to go beyond five messages, but unfortunately we have not been able to find a systematic way to prove the converse for general $m$.
\end{rem}

\subsection{Converse for the PICOD$(1)$ with \canth}
\label{subsec:converse-hyper}
The reader can find a refresher on graph theory terminology in Section~\ref{sub:graph_prelimiary}.
The critical complete--$\{s\}$ PICOD$(\cardi)$ 
we solved 
has a \nth\ which is the dual hypergraph of the complete $(m-s)$--uniform hypergraph.
Here we solve the PICOD$(1)$ whose \nth\ is a special hypergraph, namely, a circular-arc hypergraph. 

\begin{thm}
	\label{thm:circular-arc_optimality}
	For a PICOD$(1)$ with $m$ messages and with \canth, the optimal code length satisfies $\ell^*\leq 2$.
	In particular, the optimal number of transmissions is $\ell^* =2$ unless the \nth\ is a $1$-factor hypergraph.
\end{thm}
The proof can be found in Section~\ref{sec:picod_with_special_network_topology_hypergraph}.

\begin{rem}
The achievability part of Theorem~\ref{thm:circular-arc_optimality} is based on the following property of a \cah: if two vertices belong to an edge, then all vertices (cyclic) between these two vertices must belong to the same edge.
The converse part of Theorem~\ref{thm:circular-arc_optimality}, which is in Proposition~\ref{prop:l=1_condition}, is proved by showing that there exists a user that can decode one more message other than its desired message if a $1$-factor does not exist.
By showing the existence of such a user, regardless of the choices of desired messages and code sent by the transmitter, we obtain a tight lower bound on the optimal code length.
\end{rem}

The proofs of the converse results summarized in this section will be given in the following sections.

\section{Layer Counting Converse: Proof of Theorem~\ref{thm:layer_cont_opt}} 
\label{sec:layer_counting_converse}
Recall that the complete--$S$ PICOD$(\cardi)$, for a given set $S\subseteq[0:m-\cardi]$, comprises $n=\sum_{s\in S}\binom{m}{s}$ users where the side information sets are all possible distinct subsets of size $s$ of $m$ messages, for all $s\in S$.
The proof of Theorem~\ref{thm:layer_cont_opt} relies on idea of \emph{decoding chain}, which gives a high level explanation of the proof of the following lemma (see discussion after the proof), namely, the number of messages decoded along this chain provides a converse on $\ell^*$.

\begin{lemma}
	\label{lem:converse_on_decodable_msg}
	In a PICOD$(\cardi)$ with $m$ messages and $n$ users, for any ordering of the users (i.e., up to relabeling the users) we have 
	\begin{align}
	\ell^*\geq \sum_{i=1}^{n}\left| D_i\setminus \cup_{j=1}^{i-1}(A_j\cup D_j)\right|.
	\label{eq:converse_on_decodable_msg}
	\end{align}
\end{lemma}
\begin{IEEEproof}[Proof of Lemma~\ref{lem:converse_on_decodable_msg}]
Since we have a working system, all users are satisfied by the transmission of $x^{\ell \mbit}$ of length $\ell$.
For user $u_1$ we have 
\begin{align}
H\left(W_{D_1}|x^{\ell \mbit}, W_{A_{1}}\right) \leq \ell \epsilon_\ell, 
\end{align}
where 
$\lim_{\ell\to\infty}\epsilon_\ell =0$ by Fano's inequality.
Similarly, for user $u_2$ we have
\begin{align}
H\left(W_{D_2}|x^{\ell \mbit}, W_{A_{2}}\right) \leq \ell \epsilon_\ell. 
\end{align}
Therefore we have 
\begin{align*}
	&  H\left(W_{D_1}, W_{D_2}|x^{\ell \mbit}, W_{A_{1}}, W_{A_2\setminus D_1}\right) 
  \\&= H\left(W_{D_1}|x^{\ell \mbit}, W_{A_{1}}, W_{A_2\setminus D_1}\right)
	 + H\left(W_{D_2}|x^{\ell \mbit},  W_{A_{1}}, W_{A_2\setminus D_1}, W_{D_1}\right)
  \\&= H\left(W_{D_1}|x^{\ell \mbit}, W_{A_{1}}, W_{A_2\setminus D_1}\right)
	 + H\left(W_{D_2}|x^{\ell \mbit},  W_{A_{2}}, W_{A_1\cup D_1}\right)
  \\&\leq H\left(W_{D_1}|x^{\ell \mbit}, W_{A_{1}}\right)
        + H\left(W_{D_2\setminus (A_1\cup D_1)}|x^{\ell \mbit}, W_{A_{2}}\right)
  \\&\leq 2 \ell \epsilon_\ell. 
\end{align*}
By continuing with the same reasoning, we get
\begin{align}
	H\left(W_{\cup_{i=1}^n D_i}|x^{\ell \mbit}, W_{\cup_{i=1}^n (A_{i}\setminus \cup_{j=1}^{i-1}D_j)}\right) 
	\leq n \ell \epsilon_\ell. 
\end{align}
Since the messages are independent and uniformly distributed with entropy $\mbit$ bits, 
and since the code is binary, we conclude
\begin{align*}
  &  \sum_{i=1}^{n}\left| D_i\setminus \cup_{j=1}^{i-1}(A_j\cup D_j)\right| \mbit 
\\&= \left| \cup_{i=1}^n \left(D_i\setminus \cup_{j=1}^{i-1}(A_j\cup D_j)\right)\right|\mbit
\\&= H\left(W_{ \cup_{i=1}^n \left(D_i\setminus \cup_{j=1}^{i-1}(A_j\cup D_j)\right)}\right)
\\&= H\left(W_{ \cup_{i=1}^n \left(D_i\setminus \cup_{j=1}^{i-1}(A_j\cup D_j)\right)}|W_{\cup_{i=1}^n (A_{i}\setminus \cup_{j=1}^{i-1}D_j)}\right)
\\&\leq I\left(W_{ \cup_{i=1}^n \left(D_i\setminus \cup_{j=1}^{i-1}(A_j\cup D_j)\right)}; x^{\ell \mbit} \big| W_{\cup_{i=1}^n (A_{i}\setminus \cup_{j=1}^{i-1}D_j)}\right) + n \ell \epsilon_\ell 
\\&\leq H\left(x^{\ell \mbit}|W_{\cup_{i=1}^n (A_{i}\setminus \cup_{j=1}^{i-1}D_j)}\right)+  n \ell \epsilon_\ell 
\\&\leq H(x^{\ell \mbit})  + n\ell \epsilon_\ell 
\\&\leq \ell \mbit  + n\ell \epsilon_\ell, 
\end{align*}
which implies that
\begin{align}
	\ell \geq \sum_{i=1}^{n}\left| D_i\setminus \cup_{j=1}^{i-1}(A_j\cup D_j)\right|,
	\label{eq:lemma1ends}
\end{align} 
for constant $(n,\mbit)$, sufficiently large $\ell$, and any valid codes. Therefor the bound in~\eqref{eq:lemma1ends} must hold for the optimal code length as well, thus proving~\eqref{eq:converse_on_decodable_msg}.
\end{IEEEproof}

The sequence of users $u_1,u_2, \dots , u_n$ in Lemma~\ref{lem:converse_on_decodable_msg} is the \emph{decoding chain} mentioned at the beginning of this section. 
In fact, the converse in Lemma~\ref{lem:converse_on_decodable_msg} can also be thought of as the \emph{acyclic induced subgraph} converse for the all unicast IC problem~\cite{index_coding_with_sideinfo}, where each user desires multiple messages, as opposed to a single message.
The users with $\left| D_i\setminus \cup_{j=1}^{i-1}(A_j\cup D_j)\right|\neq 0$ form an acyclic induced subgraph in the graph representation of the IC.
Therefore, in Lemma~\ref{lem:converse_on_decodable_msg} the value of $\left| \cup_{i=1}^n \left(D_i\setminus \cup_{j=1}^{i-1}(A_j\cup D_j)\right)\right|$ depends on the choice of the order for the users, that is, we can relabel the users in order to find the tighest bound provided by Lemma~\ref{lem:converse_on_decodable_msg}.
Finding such an order for Lemma~\ref{lem:converse_on_decodable_msg} illustrates the intuition for the converse proof of Theorem~\ref{thm:layer_cont_opt}: finding the user that can decode the largest number of messages.

\begin{figure}
  \centering
  \includegraphics[width=0.35\columnwidth]{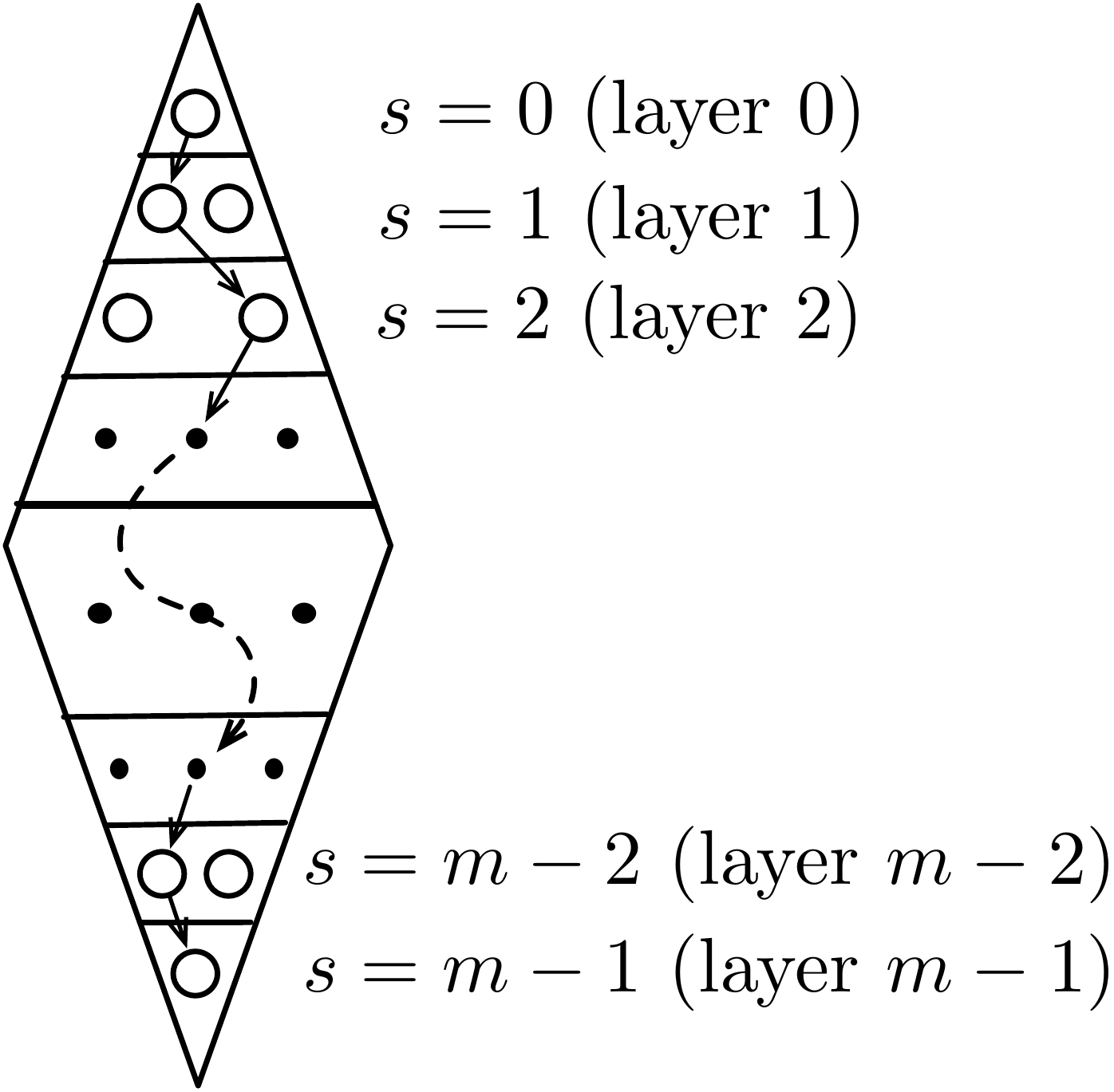}
  \caption{Layer representation of the complete--$[0:m-1]$ PICOD$(1)$ problem.}
  \label{fig:complete_setting}
\end{figure}

To illustrate the method of finding the user who can decode the largest number of messages, we introduce the \emph{layer representation} of the complete--$S$ PICOD$(\cardi)$.
As an example, the layer representation for the complete--$[0:m-1]$ PICOD$(1)$ problem is given in Fig.~\ref{fig:complete_setting}.
In Fig.~\ref{fig:complete_setting},
all the users with the same size of the side information set are said to form a layer, and there are in total $m$ layers;
the $i$-th layer contains the users whose side information set has size $i\in[0:m-1]$, and 
the number of users in the $i$-th layer is $\binom{m}{i}$. 
The key observation is that, in a working system, a user $u_i$ in $i$-th layer can decode a message $w_{d_{i}}$ it does not have in its side information set $A_{i}$. 
After that, user $u_i$ is equivalent to a user $u_{i+1}$ in the $(i+1)$-th layer whose side information is  $A_{{i+1}} = A_{i} \cup \{d_i\}$.
User $u_i$ will thus be able to decode the message $w_{{d_{i+1}}}$ that is desired by user $u_{d_{i+1}}$, in addition to its own desired message $w_{d_{i}}$.
But now user $u_i$ will have  $A_{{i+2}} = A_{i} \cup \{w_{d_{i}}, {w}_{d_{i+1}}\}$, which is the side information of a user $u_{i+2}$ in the $(i+2)$-th layer.
By continuing with the same reasoning, user $u_i$ will be able to mimic one user per layer until the last layer. 
We apply this argument to the user in the $0$-th layer (there is only one such user). 
We see that the user in the $0$-th layer is able to decode one message per layer without loss of optimality, that is, the user in the $0$-th layer decodes $m$ messages in total. 
This provides a decoding chain of length $m$. In this decoding chain each user's side information set and the desired message set form the side information set of the next user.
By having such a decoding chain, we can use Lemma~\ref{lem:converse_on_decodable_msg} to show that $\ell^*\geq m$ for the complete--$[0:m-1]$ PICOD$(1)$ problem in Fig.~\ref{fig:complete_setting}.
We use this observation, and similar ones, in the following to provide a lower bound on $\ell^*$ in terms of number of messages a user can decode, which is the main idea in all our converse proofs.

The proof of Theorem~\ref{thm:layer_cont_opt} directly follows this idea of counting the layers in a layer representation of a decoding chain. The key for the proof is the fact that each layer in the layer representation for the complement-consecutive complete--$S$, where $S=[0:m-\cardi]\setminus[\so: \st]$, contains all users with side information set of the same size. After the user has decoded its desired message(s), we can map this user to another user in a higher layer. Such a mapping forms a decoding chain, starting from the user in the $0$-th layer, provides a lower bound on $\ell^*$.

We are now ready to prove Theorem~\ref{thm:layer_cont_opt} for the complement complete--$S$ with $m$ messages and $S=[0:m-\cardi]\setminus[\so : \st]$.
\begin{IEEEproof}[Proof of Theorem~\ref{thm:layer_cont_opt}]
	Consider the PICOD$(\cardi)$ where $S=[0:m-\cardi]\backslash [\smin:\smax] = [0:\so-1]\cup[\st+1:m-\cardi]$ for some $0< \so\leq \st  < m-\cardi$. 
	We aim to find the decoding chain that has the largest number of messages/users along the chain. In each layer of the layer representation we find a user for the decoding chain. Therefore the chain contains $|S|$ users, where 
	\begin{align}
	|S| = \so + m-t-\st.
	\end{align}
	We first find $\so$ users, one user per layer for the layers indexed by $[0:\so-1]$, as done in the example in Fig.~\ref{fig:complete_setting}. 
	Then, the user $u_j|_{j=\so+1}$ is found in layer $\st+1$ such that $|A_j|_{j=\so+1}|=\st+1$; 
	we want $A_{\so+1}\supseteq A_{\so}$ so user $u_{\so+1}$ can be mimicked by user $u_{\so}$; 
	we want $|D_{\so}\cap A_{\so+1}|$ to be as large as possible so the number of messages decoded in the decoding chain can be maximized; 
	details on how this is done will be given next.
	Finally, we find other $|S|-\so-1 = m-t-\st-1$ users to complete the decoding chain, one user per layer for the layers indexed by $[\st+2:m-\cardi]$, as done in the example in Fig.~\ref{fig:complete_setting}.
	
	Assume all users are satisfied by the transmission of $x^{\ell \mbit}$.
	Let $u_1$ be the user with empty side information set, i.e., $A_1=\emptyset$. 
	Since all users are satisfied, $u_1$ can decode at least one message not in its side information set; 
	denote the index of such a message as $d_{1,1}$.
	Layer~$1$ contains the users with side information set of size $1$.
	There exists a user in layer~$1$, say $u_2$, with side information $A_2=A_1\cup\{d_{1,1}\} = \{d_{1,1}\}$ and desired message $d_{2,1}\notin A_2$.
	By continuing with this reasoning we can find users up to user $u_{\so}$: user $u_{\so}$ has side information set $A_{\so} = A_{\so-1} \cup \{d_{\so-1,1}\}= \{d_{1,1}, \ldots, d_{\so-1,1}\}$ and desires message $d_{\so,1}\not\in A_{\so}$.
	We would be tempted to say that the next user in the decoding chain should be user $u_{\so+1}$ with side information set $A_{\so} \cup D_{\so}$; however $|A_{\so} \cup D_{\so}| =  |A_{\so}|+|D_{\so}| = \so-1+t$ may be strictly less than the size of the side information of the next layer of users present in the systems, which is $\st+1$. 
	For this reason, user $u_{\so+1}$ is chosen in layer $\st+1$ as follows:
	if $\so-1+t < \st+1$, choose any user in layer $\st+1$ as $u_{\so+1}$ such that $A_{\so+1} \supset A_{\so} \cup D_{\so}$, i.e., the user that $u_{\so}$ can mimic by providing a genie side information $A_{\so+1} \setminus (A_{\so} \cup D_{\so})$;
	otherwise we choose $u_{\so+1}$ to be the user with $A_{\so+1} \subseteq A_{\so} \cup D_{\so}$, i.e., the user that $u_{\so}$ can mimic.
	From this point onwards, the next users in the decoding chain can again be chosen such that $A_j=A_{j-1}\cup \{d_{j-1,1}\}, \ j\in[\so+2:|S|]$. 

	Note that in the decoding chain we have $A_j=A_{j-1}\cup \{d_{j-1,1}\}$ for $j\in[|S|]\setminus\{1,\so+1\}$, 
	$A_1=\emptyset$, and $A_{\so+1}\supset A_{\so}$.
	These users satisfy $A_j\supset A_{j-1}$, for all $j\in[2:|S|]$. 
	Therefore we have $\left| D_i\setminus \cup_{j=1}^{i-1}(A_j\cup D_j)\right|\geq 1$ for all $i\in [|S|]\setminus \{1, \so+1\}$, 
	$\left| D_i \right|_{i=1}=\cardi$, and 
	$\left| D_i\setminus \cup_{j=1}^{i-1}(A_j\cup D_j)\right|_{i=\so+1}=\min\{\cardi, \st+1+\cardi-(\so-1+t)\}=\min\{\cardi, \st-\so+2\}$. 
Therefore, by Lemma~\ref{lem:converse_on_decodable_msg} we have
	\begin{align*}
	\ell^*
	  &\geq \sum_{i\in[|S|]}\left| D_i\setminus \cup_{j=1}^{i-1}(A_j\cup D_j)\right|
	\\&\geq \cardi+(\so-1)+\min\{t,\st-\so+2\}+(m-t-\st-1) 
	\\&\geq m+\so-\st-2 + \min\{\cardi,\st-\so+2\} 
	\\&=\min\{m+\cardi+\so-\st-2, m\}.
	\end{align*}

The value $\ell^* = \min\{m+\cardi+\so-\st-2, m\}$ can be achieved as follows.
	By the scheme in Proposition~\ref{prop:achievability_complete-s} with partition $S = S_1 \cup S_2$, with $S_1:=[0:\so-1]$ and $S_2:=[\st+1:m-\cardi]$, all users in group $S_1$ are satisfied with $(\so-1)+t$ transmissions, and all users in group $S_2$ are satisfied with $m-(\st+1)$ transmissions; therefore, we have a code of length $m+\so+t-\st-2$.
	Also, we can always transmit all $m$ messages one by one, resulting in a code of length $m$. 
	Therefore, we can achieve the lower bound by using the code among the above two with the shortest length.

	This concludes the proof of Theorem~\ref{thm:layer_cont_opt}.
\end{IEEEproof}

\begin{rem}
\label{rem:layer counting not tight}
The proof of Theorem~\ref{thm:layer_cont_opt} constructively builds a decoding chain.  
The decoding chain starts from the user in the lowest layer.
The next user in the chain is chosen in the next layer, based on the side information and desired message of the previous one.
The chain ends at the highest layer.
However, this construction, where each layer contributes at most one user to the decoding chain, is not always tight.

As shown in~\cite{itw_2017}, for the complete--$S$ PICOD$(1)$ where $S=[1:q]$ or $S=[q:m-2]$, $1\leq q\leq m-2$, the optimal code length is $\ell^*=|S|+1$. In other words, there exists a decoding chain which includes two users with the same size of side information, where one of the users can mimic the other one. 

The proof in~\cite{itw_2017} is a case-by-case reasoning, where the different cases are for different choices of desired messages of the users. For the complete--$S$ PICOD$(1)$ for general $S=[\so: \st]$, the number of cases becomes too large to be tractable. Thus a method that does not relay on a case-by-case study becomes necessary. This is what we are going to do in the next section.
The two cases considered in~\cite{itw_2017} are special cases of Theorem~\ref{thm:complete_consecutive} proved next.
\end{rem}

\section{Critical Case: complete--$\{s\}$ PICOD$(\cardi)$ with $m=2s+\cardi$ messages} 
\label{sec:complete_s_picod_m=2s+1}

To overcome the limitation of the case-by-case reasoning highlighted in Remark~\ref{rem:layer counting not tight}, 
we shall turn to an \emph{existence proof} technique for Theorem~\ref{thm:complete_consecutive}.
Loosely speaking, when dealing with general consecutive complete--$S$ PICOD$(\cardi)$ with $S=[\so : \st]$, we treat all users and all the various desired message assignments at once.
Before we prove Theorem~\ref{thm:complete_consecutive} in full generality, we consider the critical case in~\eqref{eq:critical}. 
We shall see that all other consecutive complete--$S$ cases can be deduced from the critical one. 
Therefore, this section contains the proof for the following key result:
\begin{prop}
\label{prop:power_user_exist}(The critical case)
For the complete--$\{s\}$ PICOD$(\cardi)$ with $m=2s+\cardi$ messages, the optimal code length is $\ell^*= s+\cardi$. Specifically, given a valid code, there always exists a user that can decode $\ell^*=s+\cardi$ messages.
\end{prop}

As for the layer counting converse used in Theorem~\ref{thm:layer_cont_opt}, we shall show that under the assumption that all users can decode at least one message outside their side information set, there must exists a user that can mimic other users and decodes $\ell^*= s+\cardi$ messages regardless of the desired messages of all the users. 
Note that in the complete--$S$ PICOD$(\cardi)$ where $|S|=1$, only one layer exists in the layer representation. 
Thus by the constructive method in Theorem~\ref{thm:layer_cont_opt}, we only obtain the trivial bound $\ell^*\geq 1$.
However, we do need to find the specific user that can decode $\ell^*= s+\cardi$ messages, but only show its existence.
So we turn to an existence proof, which is largely based on combinatorics ideas.
Specifically, for all possible desired message set assignments for the users, given a valid code that satisfies all users, we show that there exists a user that can decode $\ell^*= s+\cardi$ messages. 
We start by introducing next the two main ingredients needed in the proof of Proposition~\ref{prop:power_user_exist}.

\subsection{Proposition~\ref{prop:power_user_exist}: Converse Main Ingredient~1: Block Cover} 
\label{ssub:block_cover_for_the_users}

So far we used the idea of decoding chain to show that a user can decode more than its set of desired messages.
The decoding chain depends on the choice of desired messages at the users. 
Once the desired messages change, the decoding chain may change as well.
Here we are only interested in the existence of such a decoding chain of a given length.
In other words, we show the existence of a decoding chain of a certain length regardless of the choice of desired messages at the users. We start with a simple example to showcase a problem we faced when 
considering different message assignments.

\begin{example}\label{ex:s=1,m=3}
Consider the complete--$\{1\}$ PICOD$(1)$, i.e., $s=\cardi=1$, 
with $m=2s+1=3$ messages for which $\ell^\star=s+1=2$ is the smallest number of transmissions needed to satisfy all the $n=\binom{m}{s}=3$ users. 
Say that
$u_1$ knows $A_1=\{1\}$ and desires $d_1=2$;
$u_2$ knows $A_2=\{2\}$ and desires $d_2=1$; and
$u_3$ knows $A_3=\{3\}$ and desires $d_3=1$.
By sending $w_1$, users $u_2$ and $u_3$ are satisfied;
by sending $w_2$, user $u_1$ is satisfied.
By the decoding chain argument, user $u_3$ is able to mimic $u_1$ (because he decodes the message that is the side information set of user $u_1$)
and therefore can also decode $w_2$; on the contrary,
users $u_2$ and $u_3$ can not decode any more messages other than the desired one.
However, another choice of desired messages can be $d_1=3, d_2=1, d_3=1$; with this, users $u_1$ and user $u_3$ can only decode their desired messages while user $u_2$ can mimic user $u_1$ thus is able to decode two messages.
\end{example}

As Example~\ref{ex:s=1,m=3} shows for the case $\cardi=1$, 
for a specific user, there is always an optimal choice of desired messages such that this user cannot decode any message other the desired one.
However, we also note that for any choice of desired messages, there always exists a user that can decode two messages.
In the critical case setting, we shall prove that regardless of the choice of desired messages, there always exists a user who can decode $s+\cardi$ messages.
Since there are $\binom{s+\cardi}{\cardi}^{\binom{2s+\cardi}{s}}$ (doubly exponential in $s$) possible choices of desired messages, finding explicitly such a user for every case is intractable.
Therefore, our converse shows the existence of such a user. 
The main idea of the existence proof is as follows.

Instead of checking all possible different choices of desired message sets at the users, we reason on the size of the decoding chain for that user. 
By assumption, every user can decode $\cardi$ messages outside its side information set. 
Some users may be able to decode more messages because they can mimic other users. 
After receiving a valid code, we aim to show that every user eventually knows at least $s+\cardi$ messages, including the $s$ messages in its side information set and the (at least) $\cardi$ decoded ones.
Say that user $u_j$, with side information $A_j$, eventually can decode the messages indexed by $B_j\supseteq D_j$.  
One can think of the set $C_j:=A_j\cup B_j$ as a \emph{block} that \emph{covers} the side information set $A_j$, by which we mean that the set $C_j$ is a proper superset of $A_j$.
User $u_j$ can also mimic any users $u_k$ whose side information set satisfies $A_k\subset C_j$. 
Therefore the desired message set for all the users $u_k$ whose side information $A_k\subset C_j$ is $D_k\subset C_j$.
For any subset of users we can find a collection $\mathcal{C}$ such that, for every side information set $A_j$, there is a cover $C_j\in \mathcal{C}$ such that $C_j= A_j\cup B_j$ where $B_j$ is the largest set of the messages that user $u_j$ can decode.
By this definition, this \emph{block cover} / collection $\mathcal{C}$ satisfies the following properties:
\begin{enumerate}
	\item\label{item:BlockCover-P1} \ [BlockCover-P\ref{item:BlockCover-P1}] 
	For every $s$-element subset of $[m]$, there exists at least one $C\in \mathcal{C}$ that contains this subset.
	\item\label{item:BlockCover-P2} \ [BlockCover-P\ref{item:BlockCover-P2}]
	$s < |C| \leq m$ for all $C\in \mathcal{C}$.
	\item\label{item:BlockCover-P3} \ [BlockCover-P\ref{item:BlockCover-P3}]
	For all $P\subseteq [|\mathcal{C}|]$, we have $|\cap_{j\in P} C_j| \notin [s:s+\cardi-1]$.
\end{enumerate}
\begin{IEEEproof}[Proof of BlockCover Properties]
Properties BlockCover-P\ref{item:BlockCover-P1} and BlockCover-P\ref{item:BlockCover-P2} follow by the definition of block cover, while property BlockCover-P\ref{item:BlockCover-P3} holds because if we have $|\cap_{j\in P} C_j| \in [s:s+\cardi-1]$ for some $P\subseteq [|\mathcal{C}|]$, we can have a user with side information set $A^\prime \subseteq \cap_{j\in P} C_j$ with corresponding decoding set $D^\prime$ and this leads to the following contradiction.
By definition of intersection $A^\prime\subset C_j, \ \forall j\in P$; 
but also by definition of block cover $D^\prime\subset C_j, \ \forall j\in P$;
thus $A^\prime \cup D^\prime\subseteq C_j, \ \forall j\in P$, which implies $|\cap_{j\in P} C_j| \geq |A^\prime \cup D^\prime| = |A^\prime|  + |D^\prime| \geq s +\cardi$ that contradicts the starting assumption $|\cap_{j\in P} C_j| \in [s: s+\cardi-1]$.
\end{IEEEproof}

This block cover idea was inspired by the so-called \emph{generalized Steiner system} in combinatorial design~\cite{generalized_steiner_system}.
An $\mathcal{S}(s,*,m)$ generalized Steiner system consists of blocks / sets such that each subset of size $s$ from the ground set of size $m$ is covered exactly once. 
In a critical PICOD$(\cardi)$ setting, the collection of blocks $\mathcal{C}$ also covers all $s$-element subsets of $[m]$ (i.e., all users' side information sets). 
But our problem is not exactly a generalized Steiner system because 
an $s$-element subset may be contained in more than one block as long as it is not an exact intersection of the blocks--see Property BlockCover-P\ref{item:BlockCover-P3}.
Therefore, our block cover can be seen as a relaxed generalized Steiner system.
To the best of our knowledge no results are available for this specific relaxed generalized Steiner system.

For the critical case we aim to show that there is a user who can decode $s+\cardi$ messages (as in Example~\ref{ex:s=1,m=3}).
We argue it by contradiction. 
Assume no user can decode $s+\cardi$ messages,
that is, every user can decode at least $\cardi$ and at most $s+\cardi-1$ messages by mimicking other users.
In terms of block cover, this indicates that we can have a block cover $\mathcal{C}$ with $\max_{C\in \mathcal{C}}\{|C|\}\leq (s+\cardi-1)+s <m=2s+\cardi$.
Our argument of showing that there always exists a user that can decode $\cardi+s$ messages for the critical case is equivalent to showing that a block cover with size at most $2s+\cardi-1$ cannot exist.
Our combinatorial proof shows that the existence of a choice of desired messages such that $\cardi+s \leq |C_j|\leq 2s+\cardi-1, \forall j\in[|\mathcal{C}|]$ leads to the existence of a user that can decode $\cardi+s$ messages, thus $\max|C_j|=2s+\cardi$, which is a contradiction.
Therefore must exists a user whose block cover has size $m=2s+\cardi$.

\subsection{Proposition~\ref{prop:power_user_exist}: Converse Main Ingredient~2: Maximum Acyclic Induced Subgraph (MAIS) Bound} 
\label{ssub:mais_bound}

Recall that for a PICOD$(\cardi)$, each user chooses $\cardi$ desired messages outside its side information set.
The collection of the desired message sets for all the users users is denoted as $\mathcal{D}=\{D_1,\dots,D_n\}$, where $n={2s+\cardi\choose s}$. 
Once $D$ is chosen, the PICOD$(\cardi)$ reduces to a \emph{multi-cast IC} where each user requests $\cardi$ messages;
we can make one user to be $\cardi$ users with the same side information sets but each with a distinct single desired message;
the multi-cast IC with $n$ users becomes a multi-cast IC with $tn$ users, each requesting one message.

Similarly to the classic all-unicast IC, we can represent in a directed graph / digraph the side information sets and the desired messages of a multi-cast IC where each user desires a single message~\cite{index_coding_with_sideinfo}.
Pick a subset $U\subseteq[tn]$ of users who desire different messages and create a digraph $\mathsf{G}(U)$ 
as follows.
The vertices $V(\mathsf{G})\subseteq W$ represent the desired messages by the users in $U$. 
A directed arc $(w_i,w_j)\in E(\mathsf{G})$ exists if and only if the user who desires $w_i$ has $w_j$ in its side information set. 
$\mathsf{G}$ is called \emph{acyclic} if it does not contain a directed cycle.
The \emph{size} of $\mathsf{G}$ is the number of the vertices in, i.e. $|V(\mathsf{G})|=|U|$. 
For the all-unicast IC, the maximum size of $U$ such that the corresponding digraph $\mathsf{G}(U)$ is acyclic serves as a converse bound on the optimal code length. This converse is known as \emph{maximum acyclic induced subgraph} (MAIS) bound~\cite{index_coding_with_sideinfo}.

For the PICOD$(t)$, a similar \text{MAIS} bound can be found, which is the maximum size of the acyclic digraph $\mathsf{G}(U)$ created by the choice of users $U\subseteq [tn]$ such that they all desire different messages.
Since \text{MAIS} depends on the desired message set $\mathcal{D}$, we denote its size as $|\text{MAIS}(\mathcal{D})|$.
Thus, for the PICOD$(\cardi)$ as for multi-cast IC, the size of \text{MAIS} is a converse bound on $\ell$~\cite{index_coding_with_sideinfo}, namely, $\ell\geq |\text{MAIS}(\mathcal{D})|$.

Finding the \text{MAIS} for the all-unicast IC is known to be an NP-hard problem~\cite{21np-hard-problems} in general.
Finding the \text{MAIS} for the multi-cast IC appears to be more difficult since one needs to check every possible choice of users with distinct desired messages. 
Finding the \text{MAIS} for the PICOD$(\cardi)$ problem seems even more complicated since each choice of $\mathcal{D}$ in the PICOD$(\cardi)$ corresponds to a multi-cast IC, and in addition one needs to find the best $\mathcal{D}$ in terms of code length. 
Therefore, finding the \text{MAIS}  for the PICOD$(\cardi)$ by solving all possible all-unicast IC problems appears intractable.
Therefore, our existence proof does not find the exact \text{MAIS} for the PICOD$(\cardi)$, but only bounds on its size, i.e., $\max_{\mathcal{D}}|\text{MAIS}(\mathcal{D})|$.
Towards this goal, we have the following properties: 
\begin{enumerate}
	\item\label{item:MAIS-P1} \ [MAIS-P\ref{item:MAIS-P1}]
	for the critical complete--$\{s\}$ PICOD$(\cardi)$ with $m=2s+\cardi$ messages,
	$|{\text{MAIS}}(\mathcal{D})|=s+\cardi$ for certain $\mathcal{D}$ if and only if there exists a user who decodes $s+\cardi$ messages.
	\item\label{item:MAIS-P2} \ [MAIS-P\ref{item:MAIS-P2}]
	for the critical complete--$\{s\}$ PICOD$(\cardi)$ with $m=2s+\cardi$ messages,
	if there exists a $\mathcal{D}$ such that $|\text{MAIS}(\mathcal{D})|\leq s+\cardi-1$, there exists a $\mathcal{D}^\prime$ where  $|\text{MAIS}(\mathcal{D}^\prime)|=s+t-1$.
\end{enumerate}

\begin{IEEEproof}[Proof of Property \text{MAIS}-P\ref{item:MAIS-P1}]
	On the one hand, if $|\text{MAIS}(\mathcal{D})|=s+\cardi$, there are $s+\cardi$ users who desire different messages. These users form an acyclic induced subgraph. 
	We can obtain a decoding chain from the acyclic induced subgraph, in which the first user has side information of all $s$ messages that are not desired by these $s+\cardi$ users. 
	The first user, by decoding its desired message, can mimic all the other users and 
	eventually decode $s+\cardi$ messages. 

	On the other hand, if there is one user who can decode $s+\cardi$ messages, 
	there are $s+\cardi-1$ users that can be mimicked by it with different desired messages.
	These $s+\cardi$ users form an acyclic induced subgraph of size $s+\cardi$.
	Then $|\text{MAIS}(D)|=s+\cardi$. 
\end{IEEEproof}

\begin{IEEEproof}[Proof of Property \text{MAIS}-P\ref{item:MAIS-P2}]
We prove the claim by showing that for a choice of desired messages $\mathcal{D}$ that has a $|\mais(\mathcal{D})|=a$ for some integer $a<s+\cardi$, we can always find another choice of desired messages $\mathcal{D}^\prime$ such that $|\mais(\mathcal{D}^\prime)|=a+1$.

Assume there exists a $\mathcal{D}$ such that, for some integer $a<s+\cardi$, satisfies $|\mais(\mathcal{D})|=a$. 
	For this $\mathcal{D}$, the PICOD$(\cardi)$ can be seem as a unicast IC with $[\cardi n]$ users, whose graph representation has an induced acyclic subgraph of size $a$ and all induced subgraphs of size strictly larger than $a$ are cyclic.
	Without loss of generality, let $\{u_1, \dots, u_a\}$ be the set of users that form this MAIS who have desired messages $\{w_{d_1}, \dots, w_{d_a}\}=[a]$. 
	By the definition of MAIS, any user with side information $A \subseteq [a+1: m]$ must have desired message $d\in [a]$; this is so because any user with $A \subseteq [a+1: m]$ and $d\in [a+1:m]$ can be added to the users $u_1, \dots, u_a$ to form an acyclic subgraph of size $a+1$, which would contradict to the assumption that $|\mais|=a$.
	 
Based on $\mathcal{D}$ we construct $\mathcal{D}^\prime$ such that $|\mais(\mathcal{D}^\prime)|=a+1$ as  follows.
	Choose a user $u^\prime$ with side information $A^\prime \subseteq [a+1: m]$ and change its desired message to $d^\prime \in [a+1:m]\setminus A^\prime$ (it was $d^\prime\in [a]$ in $\mathcal{D}$).
	Since $a<s+t$ we have $|[a+1:m]|\geq s+1$ and $|[a+1:m]\setminus A^\prime|\geq 1$, thus such a user $u^\prime$ and its desired message $d^\prime$ can be found. Moreover, by construction the users in $\{u_1,\dots, u_a, u^\prime\}$ form an acyclic subgraph of size $a+1$. 

Next, we show that any induced subgraph of size strictly larger than $a+1$ in the IC represented by $\mathcal{D}^\prime$ is cyclic. 
This can be seen as follows. Note that from $\mathcal{D}$ to $\mathcal{D}^\prime$ only the desired message of $u^\prime$ was changed, therefore any induced subgraph in the IC represented by $\mathcal{D}^\prime$ that does not have $u^\prime$ also exists in the IC represented by $\mathcal{D}$. 
For any induced subgraph in the IC represented by $\mathcal{D}^\prime$ with size strictly larger than $a+1$, if it does not contain $u^\prime$, this induced subgraph exists in the IC represented by $\mathcal{D}$. By the condition $|\mais(\mathcal{D})|=a$ we know that this subgraph is cyclic.
If the induced subgraph contains $u^\prime$, remove $u^\prime$ so as to obtain an induced subgraph of size strictly larger than $a$. This newly obtained subgraph exists in the IC represented by $\mathcal{D}$. Similarly the subgraph is cyclic by $|\mais(\mathcal{D})|=a$ thus the original subgraph which contains $u^\prime$ is also cyclic. 
	This concludes that $|\mais(\mathcal{D}^\prime)|=a+1$. 

We show that we can always construct $|\mais(\mathcal{D}^\prime)|=a+1$ based on $|\mais(\mathcal{D})|=a<s+\cardi$. 
	Therefore if there exists a $\mathcal{D}$ such that $|\mais(\mathcal{D})|<s+t$, by the construction we have have a $\mathcal{D}^\prime$ such that $|\mais(\mathcal{D}^\prime)|=s+t-1$.
\end{IEEEproof}

We are now ready to prove  Proposition~\ref{prop:power_user_exist}.

\subsection{Proof of Proposition~\ref{prop:power_user_exist}} 
\label{sec:power_user_exist}

Our proof for Proposition~\ref{prop:power_user_exist} is by contradiction. Specifically, we prove that, under the assumption that there exists $\mathcal{D}^\prime$ such that $|\mais(\mathcal{D}^\prime)|=s+\cardi-1$ (see Property \text{MAIS}-P\ref{item:MAIS-P2}) and given a valid code, there must exist a user that can decode $s+\cardi$ messages. This however contradicts Property \text{MAIS}-P\ref{item:MAIS-P1}. Therefore $\mathcal{D}^\prime$ does not exist, which implies that there must exists a user that can decode $s+\cardi$ messages and $|\text{MAIS}(\mathcal{D})|=s+\cardi$ for all $\mathcal{D}$. This proves that for the critical case the optimal number of transmission is $\ell^*= s+\cardi$.

Specifically, the assumption that $|\text{MAIS}(D^\prime)|=s+\cardi-1$ implies that one can find a set of $s+\cardi-1$ users, denoted by $V$, who desire different messages and with a strict partial order on $V$ given by: for distinct $i,j\in V$, if $i<j$ then $d_j\notin A_i$.  Without loss of generality, let $[s+2:2s+t]$ be the set of the distinct $s+\cardi-1$ desired messages by the users in $V$. By the definition of \text{MAIS}, there is a user in $V$ such that its side information set satisfies $A\subset [s+1]$. This is the user that has no incoming edges in the induced acyclic subgraph of the \text{MAIS}.  Thus, a user with side information including the messages in $[s+1]$ (these messages are not desired by the users in $V$) is able to decode all the messages in $[s+2:2s+t]$.

	\begin{figure}
	  	\centering
	  	\includegraphics[width=0.6\columnwidth]{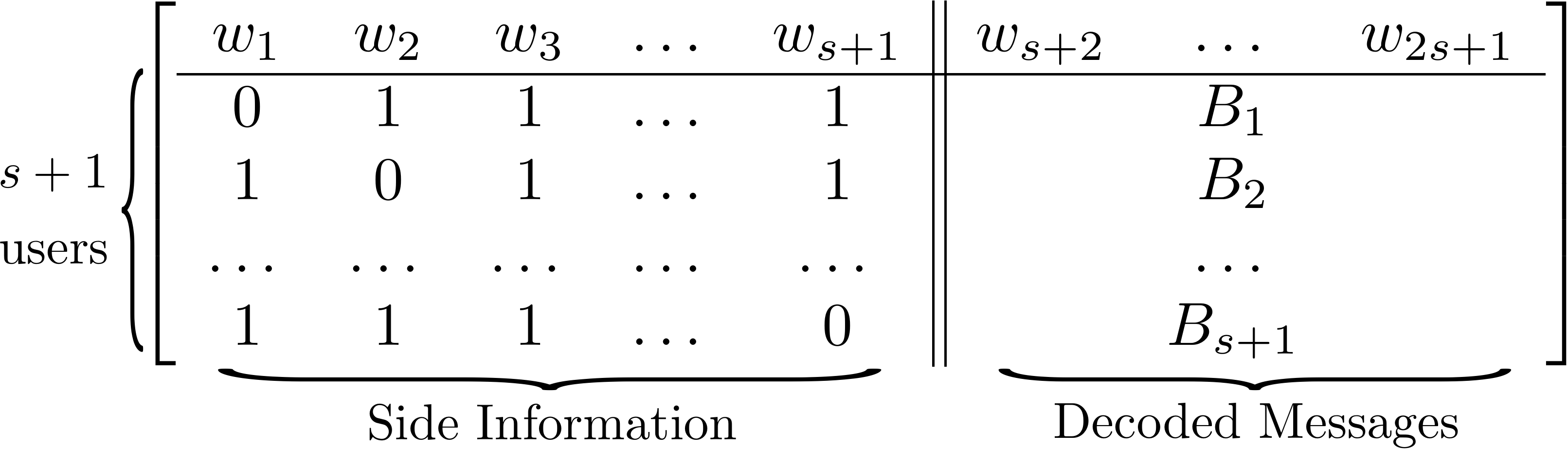}
	  	\caption{Side information sets and decoded messages for $s+1$ users for Proposition \ref{prop:power_user_exist}.Case2.}
	  	\label{fig:block_cover_critical}
	\end{figure}	
Consider the following $s+1$ users: for $i\in [s+1]$ user $u_i$ has side information $A_i = [s+1]\setminus\{i\}$. The side information sets and decoded messages of these users are illustrated in Fig.~\ref{fig:block_cover_critical} where columns are for messages and rows for users; a $0$ (resp. $1$) entry indicates the absence (resp. presence) of the corresponding message in the side information set of the user. We have one of two cases:

\emph{Proposition~\ref{prop:power_user_exist}.Case1}:
\label{emph:case_a}
Assume that for some $k\in[s+1]$ we have $B_k\cap[s+1]=[s+1]\backslash A_k$. Recall $B_k$ is the set of messages that user $u_k$ can decode and $A_k$ its side information, this user will gain the knowledge of all messages  $W_{[s+1]}$. 
It therefore can decode all the remaining messages $W_{[s+2:2s+t]}$. 
Eventually this user decodes $s+\cardi$ messages, therefore $C_k=[2s+t]$.

\emph{Proposition~\ref{prop:power_user_exist}.Case2}:
\label{emph:case_b}
For every user $i\in[s+1]$, we have $B_i\subseteq [s+2:2s+t]$--as shown in Fig.~\ref{fig:block_cover_critical}, where the side information and decodable message sets are represented by the rows of the matrx.
The left part of the matrix indicates the side information of the users, where $0,1$ entries show the absence and existence of the corresponding messages in the side information. 
%
%
	By assumption $B_i\subseteq [s+2:2s+\cardi]$ contains the indices of the messages decoded by user $u_i$ and
%
	property BlockCover-P\ref{item:BlockCover-P3},
	we have $|\cap_{i\in P}C_i|\notin [s:s+t-1]$ for any $P\subseteq [s+1]$.  
%
	Note that $|\cap_{i\in P}A_i|= s+1-|P|$ and $A_i\cap B_i = \emptyset$, 
	thus we have $|\cap_{i\in P}B_i|\notin [|P|-1: |P|+t-2], \forall P\subseteq [s+1]$.
%

In Proposition~\ref{prop:power_user_exist}.Case2, all $B_i, \ i\in[s+1]$ are non-empty subsets of a ground set $[s+2: 2s+t]$; by Lemma~\ref{lem:exist_intersection_s} in Appendix~\ref{app:lem:exist_intersection_s}, it is guaranteed that there is a $P$ such that $|[s+2: 2s+1]\cap(\cap_{i\in P}B_i)|=|P|-1$; therefore we have $\left|\cap_{i\in P}B_i\right|\in[|P|-1 : |P|+t-2]$ for some $P\subseteq [s+1]$, which contradicts what we just stated, thus this case in impossible.

Therefore only Proposition~\ref{prop:power_user_exist}.Case1 is possible.
This shows the existence of a user whose block cover is $[m]=[2s+\cardi]$.
This user can decode $s+\cardi$ messages.
But this contradicts the assumption that the \text{MAIS} bound is $|\mais(\mathcal{D}^\prime)|=2s+\cardi-1$.
Overall, this shows that for all possible choices of $\mathcal{D}$ one must have $|\text{MAIS}(\mathcal{D})|=2s+\cardi$, which implies $\ell^*\geq s+\cardi$. 
This, with the achievability in Proposition~\ref{prop:achievability_complete-s}, concludes the proof of Proposition~\ref{prop:power_user_exist}.

\subsection{Complete--$S$ where $|S|=1$} 
\label{sub:complete_s_cardinality_one}

With Proposition~\ref{prop:power_user_exist}, we can prove a more general case.
\begin{prop} (The case $|S|=1$.)
	\label{prop:card=1_complete-s}
	For the complete--$\{s\}$ PICOD$(\cardi)$ with $m$ messages, the optimal code length is $\ell^*= \min\{s+t,m-s\}$.	
\end{prop}
\begin{IEEEproof}[Proof of Proposition~\ref{prop:card=1_complete-s}]
Proposition~\ref{prop:power_user_exist} solves the case where $S=\{s\}$ and $m=2s+\cardi$.
Therefore, in the following we study the remaining two cases: $m<2s+\cardi$ and $m>2s+\cardi$.

\paragraph*{Case $m<2s+\cardi$} 
\label{ssub:proof_of_case_small_m}
    Consider 
    an integer $\alpha \leq s$ and split 
	the $n=\binom{m}{s}$ users in the system into two categories: 
	users $u_i$ with $[\alpha] \subset A_i$, 
	and the other users.
	The users in the first category 
	do not decode any message in $[\alpha]$ (since they have all these messages in their side information set);
	these users together form a complete--$\{s-\alpha\}$ PICOD$(\cardi)$ with $m-\alpha$ messages.
	Since this complete--$\{s-\alpha\}$ PICOD$(\cardi)$ is a subset of the original complete--$\{s\}$ PICOD$(\cardi)$, its optimal number of transmissions is a lower bound on the number of transmissions in the original system.
	If we take $m-\alpha = 2(s-\alpha)+t \Longleftrightarrow \alpha = 2s+t-m > 0$ then, by Proposition~\ref{prop:power_user_exist}, the optimal number of transmissions for the complete--$\{s-\alpha\}$ PICOD$(\cardi)$ with $m-\alpha= 2(s-\alpha)+t$ messages is $(s-\alpha)+t=m-s$.
	Therefore the original complete--$\{s\}$ PICOD$(\cardi)$ 
	requires at least $m-s$ transmissions, i.e., $\ell^*\geq m-s = \min\{m-s,s+\cardi\}.$

\paragraph*{Case $m>2s+\cardi$} 
\label{ssub:proof_fo_case_large_m}
	The proof is by contradiction.
	Assume there exists a $D^\prime$ such that $|\text{MAIS}(D^\prime)|=s+t-1$ and, 
	without loss of generality, that the maximum acyclic induced subgraph is formed by users with desired messages $[s+t-1]$.
	Specifically, we have users $u_i,i\in[s+t-1]$ such that $d_i=i$ and $d_j\notin A_i$ for any $j,i\in [s], j>i$ (by the definition of \text{MAIS} and its induced partial order).

	Let $U^\prime$ index the users whose side information is a subset of $[s+t:m]$, i.e., $i\in U^\prime$ if $A_i\subset [s+t:m]$.
	Apparently $1\in U^\prime$. We distinguish the following two cases.
\\
\emph{Proposition~\ref{prop:card=1_complete-s}.($m>2s+\cardi$).Case1}: 
	If there is a user $u_i \in U^\prime$ with desired message $d_i\in [s+t:m]$, we have $d_j\notin A_\cardi$ for all $j\in[s]$. Thus users $u_i,u_1,u_2,\dots,u_{s+t-1}$ form an acyclic induced subgraph of length $s+\cardi$. 
	This contradicts to the assumption that $|\text{MAIS}(D^\prime)|=s+t-1$.
\\
\emph{Proposition~\ref{prop:card=1_complete-s}.($m>2s+\cardi$).Case2}:
	For all $i\in U^\prime$ we have $d_i\in [s]$.
	By a similar reasoning as in proof of Proposition~\ref{prop:power_user_exist}, we can show that there exists a user who can decode $s+\cardi$ messages. This again contradicts the assumption that $|\text{MAIS}(D)|=s+t-1$. 

	By combining Proposition~\ref{prop:card=1_complete-s}.($m>2s+\cardi$).Case1 and Proposition~\ref{prop:card=1_complete-s}.($m>2s+\cardi$).Case2, we conclude that $|\text{MAIS}(D)| > s$.
	By Properties~\text{MAIS}-P\ref{item:MAIS-P1} 
	and~\text{MAIS}-P\ref{item:MAIS-P2} 
	we thus have $\ell^*\geq s+\cardi = \min\{m-s,s+\cardi\}$.

The achievability follows directly the schemes in Proposition~\ref{prop:achievability_complete-s}.
Since $|S|=1$, no partition is needed.
\end{IEEEproof}

\section{Complete--$S$ PICOD$(\cardi)$ where $S$ is consecutive: Proof of Theorem~\ref{thm:complete_consecutive}} 
\label{sec:complete_s_consecutive}

With Proposition~\ref{prop:card=1_complete-s},
we are ready to prove Theorem~\ref{thm:complete_consecutive} in full generality.
We consider the following three cases.

\begin{figure}[h!]
  	\centering
\begin{subfigure}{0.3\columnwidth}
	\includegraphics[width=\columnwidth]{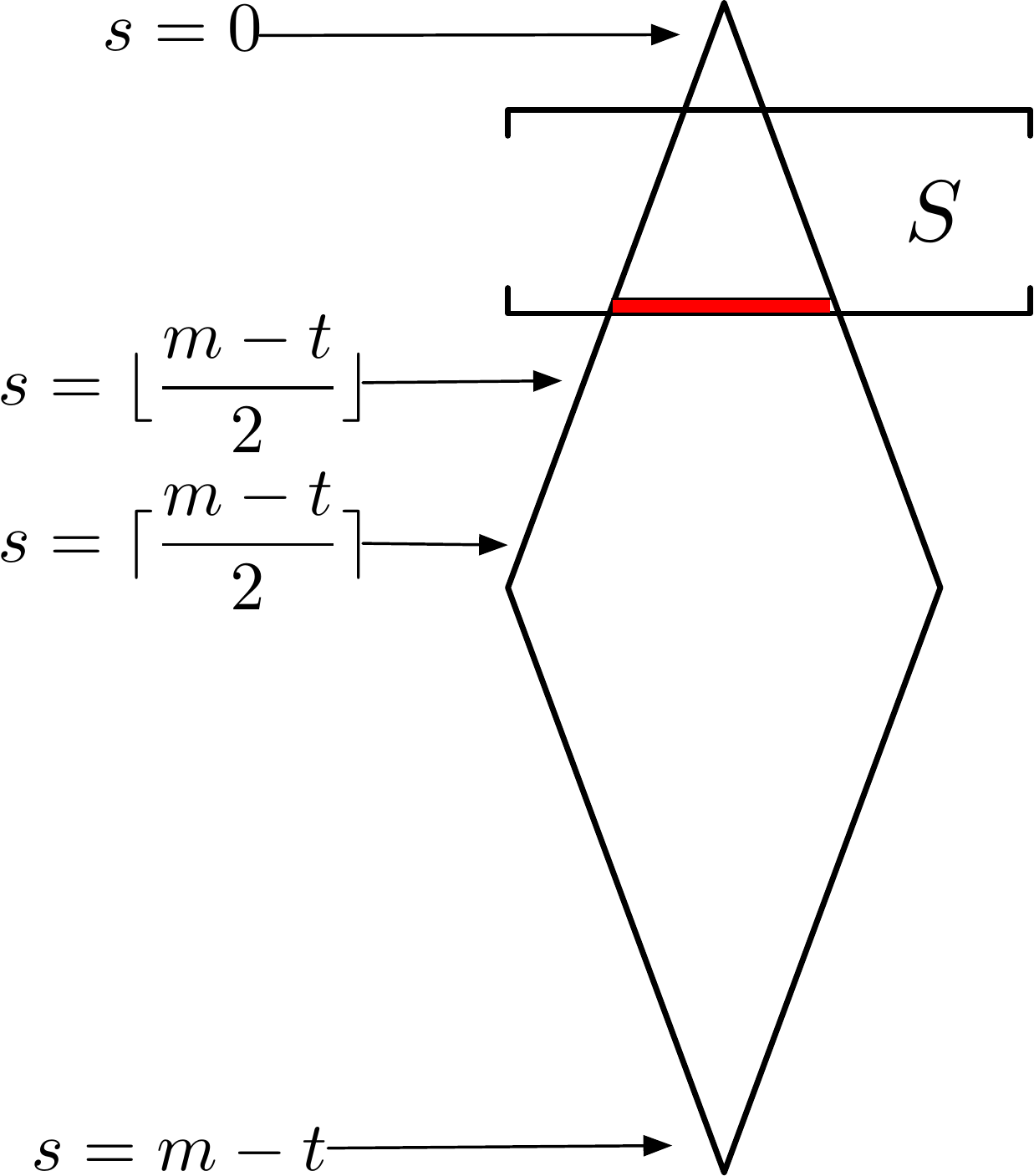}
  	\caption{$\smax \leq \ceil{(m-\cardi)/2}-1$.}
  	\label{fig:layer_representation_smax_small}
\end{subfigure}
\begin{subfigure}{0.3\columnwidth}
	\includegraphics[width=\columnwidth]{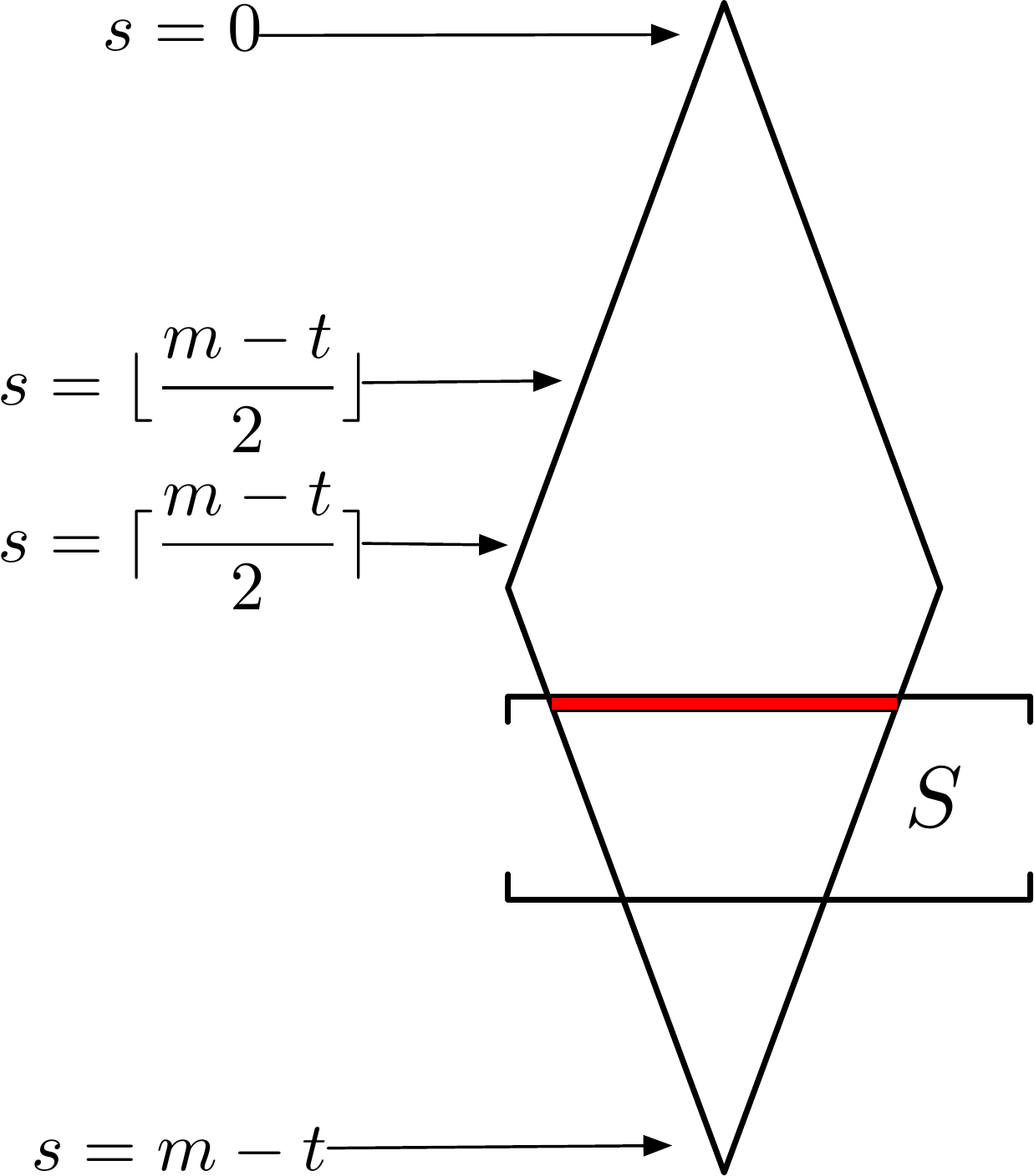}
  	\caption{$\smin \geq \floor{(m-\cardi)/2}$.}
  	\label{fig:layer_representation_smin_large}
\end{subfigure}
\begin{subfigure}{0.3\columnwidth}
	\includegraphics[width=\columnwidth]{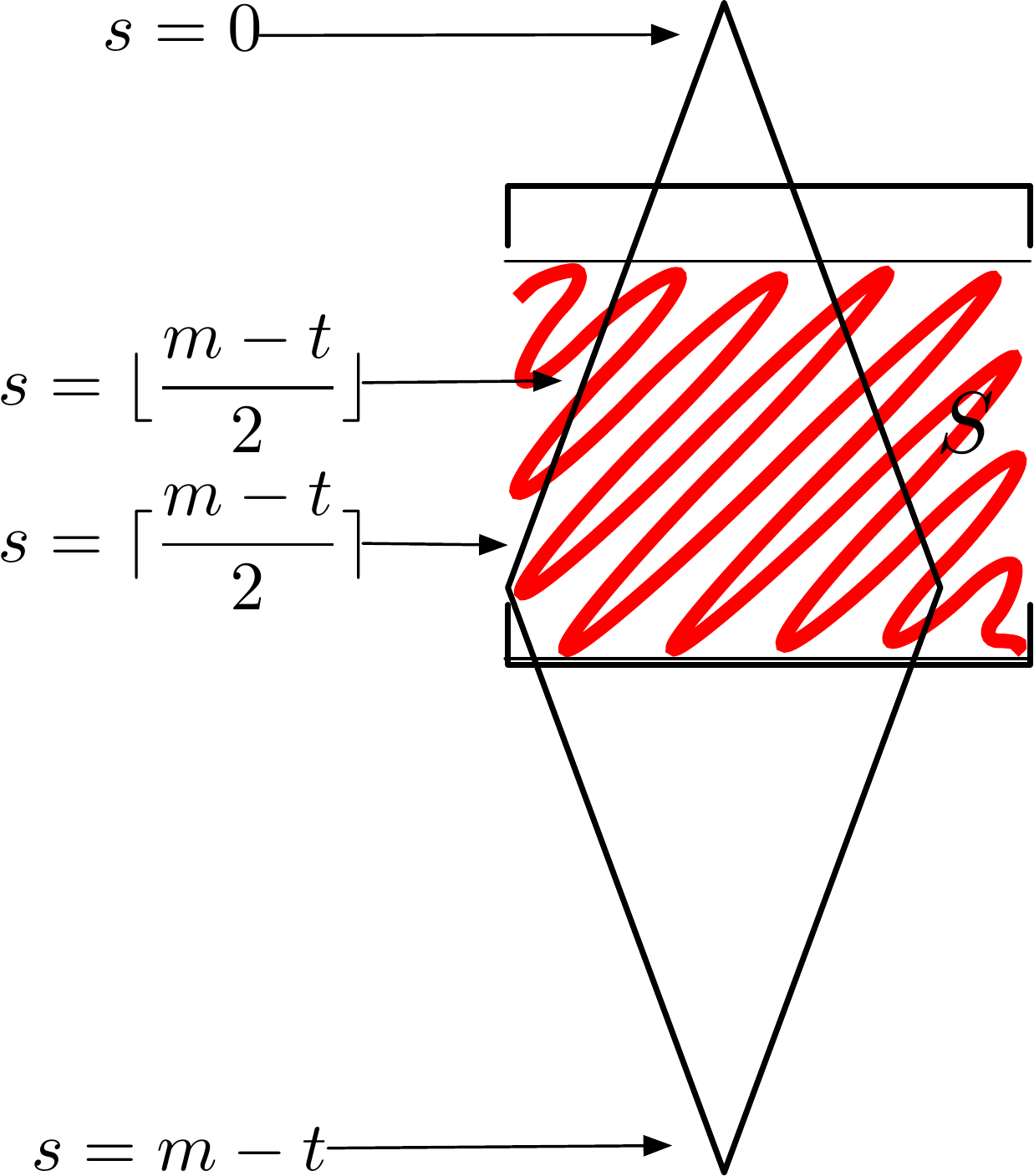}
  	\caption{$\smin \leq \floor{(m-\cardi)/2}-1 \leq \ceil{(m-\cardi)/2} \leq \smax$.}
  	\label{fig:layer_representation_middle}
\end{subfigure}
\caption{Various layer representations.}
\label{fig:LayerRepresentationForProps}
\end{figure}

\subsection{Case $\smax \leq \ceil{(m-\cardi)/2}$: $\ell^*=\smax+\cardi$} 
\label{ssub:max<m/2}
Drop all the users except those with side information set of size $\smax$, thereby obtaining a compete-$\{\smax\}$ PICOD$(\cardi)$ with $m$ messages. The layer representation of this case is shown in Fig.~\ref{fig:layer_representation_smax_small}, where the red layer is the one left after dropping users. For this system the optimal number of transmissions lower bound by $\min\{m-\smax,\smax+\cardi\} = \smax+\cardi$ (because  $\smax \leq \ceil{(m+\cardi)/2}$ in this case), which is a lower bound on the number of transmissions in the original system. By Proposition~\ref{prop:achievability_complete-s}, we have $\ell^*=\smax+\cardi$.

\subsection{Case $\smin \geq \floor{(m-\cardi)/2}$: $\ell^*=m-\smin$}
\label{ssub:min>m/2}
As for the case in Section~\ref{ssub:max<m/2}, drop all the users except those with side information of size $\smin$, thereby obtaining a compete-$\{\smin\}$ PICOD$(\cardi)$ with $m$ messages and optimal number of transmissions is $\min\{m-\smin,\smin+\cardi\} = m-\smin$ (because  $\smin \geq \floor{(m-\cardi)/2}$ in this case).  By Proposition~\ref{prop:achievability_complete-s}, we have $\ell^*=m-\smin$. The layer representation of this case is shown in Fig.~\ref{fig:layer_representation_smin_large}, where red layer is the one left after dropping users.

\subsection{Case $\smin \leq \floor{(m-\cardi)/2}-1 \leq \ceil{(m-\cardi)/2} \leq \smax$} 
\label{prop:min<m/2<max}
 	Define
	\begin{align}
	\delta   &:= \min\left\{\smax-\ceil{\frac{m-\cardi}{2}}, \floor{\frac{m-\cardi}{2}}-\smin\right\},
	\label{eq:defdelta}
	\\
	m^\prime &:= m+2\delta+\ceil{\frac{m-\cardi}{2}}-\floor{\frac{m-\cardi}{2}},
	\label{eq:defmprime}
	\\
	S^\prime &:= \left[\floor{\frac{m-\cardi}{2}}-\delta: \ceil{\frac{m-\cardi}{2}}+\delta\right].
	\label{eq:defSprime}
	\end{align}
	Drop all users except those with side information of size $s\in S^\prime$ for $S^\prime$ in~\eqref{eq:defSprime}, thereby obtaining a complete--$S^\prime$ PICOD$(\cardi)$ with $m$ messages. The layer representation of this case is shown in Fig.~\ref{fig:layer_representation_smin_large}, where red layers are the ones left after dropping users.
 	Create dummy messages $W_{[m+1: m^\prime]}$, where
	dummy messages will not be desired by any user.
To every user who was not dropped, with side information of size $s\in S^\prime$, give every $(\ceil{\frac{m-\cardi}{2}}+\delta-s)$-subset of $[m+1: m^\prime]$ as extra side information (where $\delta$ is defined in~\eqref{eq:defdelta} and $m^\prime$ in~\eqref{eq:defmprime}); 
each such user 
generates $m^\prime-m \choose \ceil{\frac{m-\cardi}{2}}+\delta-s$ new users.
 	All the users created by this procedure form a complete--$\{\ceil{\frac{m-\cardi}{2}}+\delta\}$ PICOD$(\cardi)$ with $m^\prime$ messages, whose optimal number of transmissions is 
	\begin{align*}
	&\min\left\{\ceil{\frac{m-\cardi}{2}}+\delta+t, m^\prime-(\ceil{\frac{m-\cardi}{2}}+\delta)\right\} 
  \\&=\min\left\{ \ceil{\frac{m-\cardi}{2}}+\delta+t, m+2\delta+\ceil{\frac{m-\cardi}{2}}-\floor{\frac{m-\cardi}{2}}-\ceil{\frac{m-\cardi}{2}}-\delta \right\} 
  \\&=\delta+t+\min\left\{\ceil{\frac{m-\cardi}{2}}, m-\cardi-\floor{\frac{m-\cardi}{2}}  \right\}
  \\&=\delta+t+\ceil{\frac{m-\cardi}{2}}
  \\&=\min\left\{\smax-\ceil{\frac{m-\cardi}{2}}, \floor{\frac{m-\cardi}{2}}-\smin\right\}+t+\ceil{\frac{m-\cardi}{2}}
  \\&=\min\left\{\smax+t, m-\smin\right\} 
  \\&= \ell^\prime.
	\end{align*} 
	Although the new system contains more users, any valid code for the original system works for the new one.
	Therefore the optimal code length $\ell^\prime$ is a lower bound on the optimal code length for the original system. 
	This lower bound can be attained by the scheme described in Proposition~\ref{prop:achievability_complete-s}.
	This concludes Theorem~\ref{thm:complete_consecutive}.

\section{Some other complete--$S$ PICOD$(\cardi)$} 
\label{sec:complete_nonconsecutive} 
The proofs in Section~\ref{sub:complete_s_cardinality_one} start by dropping some users in the system. 
This shows that there exists \emph{non-critical users} that do not affect the optimal code length. 
Therefore, by adding non-critical users, we can obtain a \emph{non-consecutive} complete--$S$ PICOD$(\cardi)$ where the proof used for Theorem~\ref{thm:complete_consecutive} can still provide a tight converse.

\subsection{Proof of Proposition~\ref{prop:max<m/2_nonconsecutive}} 
\label{par:max<m/2_nonconsecutive}
The converse depends only on the users with side information of size $\smax$. 
The code that satisfies the complete--$\{\smax\}$ PICOD$(\cardi)$, i.e., transmit $\smax+\cardi$ messages one at a time, also satisfies all the users with smaller size of side information.

\subsection{Proof of Proposition~\ref{prop:min>m/2_nonconsecutive}} 
\label{par:min>m/2_nonconsecutive}
The converse depends only on the users with side information of size $\smin$.
The code that satisfies the complete--$\{\smin\}$ PICOD$(\cardi)$, i.e., transmit $m-\smin$ linearly independent linear combinations of all messages, also satisfies all the users with larger size of side information.

\subsection{Proof of Proposition~\ref{prop:min<m/2<max_nonconsecutive}} 
\label{par:min<m/2<max_nonconsecutive}
The converse depends only on the users with side information of size in $[\floor{\frac{m-\cardi}{2}}-\delta: \ceil{\frac{m-\cardi}{2}}+\delta]$.
The code that satisfies the complete--$[\floor{\frac{m-\cardi}{2}}-\delta: \ceil{\frac{m-\cardi}{2}}+\delta]$ PICOD$(\cardi)$ also satisfies all the users with larger size of side information set. That is, either transmit  $\smax+\cardi$ messages one at a time, or $m-\smin$ linearly independent linear combinations of all messages.

\subsection{Proof of Proposition~\ref{prop:achi_opt_m<=5}} 
\label{sub:complete_m<=5}

Proposition~\ref{prop:achi_opt_m<=5} states that the achievable scheme in Proposition~\ref{prop:achievability_complete-s} is information theoretically optimal for the complete--$S$ PICOD$(\cardi)$ with $m\leq 5$.
The main idea behind these proofs follows the one in converse proof of Theorem~\ref{thm:layer_cont_opt}: construct a decoding chain by providing proper messages to the user as genie, in a way that the user can mimic other users and decode the desired number of messages.
Table~\ref{table:complete_s_m<=5} lists the optimal code length $\ell^*$ of all complete--$S$ PICOD$(\cardi)$ instances that are not covered by Theorem~\ref{thm:layer_cont_opt} or Propositions~\ref{prop:max<m/2_nonconsecutive},~\ref{prop:min>m/2_nonconsecutive},~\ref{prop:min<m/2<max_nonconsecutive}.

\begin{table}
\centering
\caption{Complete--$S$ PICOD$(\cardi)$ that are not covered by Theorem~\ref{thm:layer_cont_opt} or Propositions~\ref{prop:max<m/2_nonconsecutive},~\ref{prop:min>m/2_nonconsecutive},~\ref{prop:min<m/2<max_nonconsecutive}.}
\begin{tabular}{ |c|c|c|c| } 
\hline
 \multirow{2}{3em}{$m=4$}  & $S=\{0,2\}$ & $\cardi=1, 2$ & $\ell^*=\cardi+2$ \\
& $S=\{1,3\}$ & $\cardi=1$ & $\ell^*=3$ \\
\hline
\multirow{3}{3em}{$m=5$} & $S=\{0,3\}$ & $\cardi=1, 2$ & $\ell^*=\cardi+2$ \\ 
& $S=\{1,4\}$ & $\cardi=1$ & $\ell^*=3$\\ 
& $S=\{1,3\}$ & $\cardi=1, 2$ & $\ell^*=4$\\ 
& $S=\{0,1,3\}$ & $\cardi=1, 2$ & $\ell^*=4$\\
& $S=\{1,3,4\}$ & $\cardi=1$ & $\ell^*=4$\\
& $S=\{0,2,3\}$ & $\cardi=1, 2$ & $\ell^*=4$\\
& $S=\{0,2,4\}$ & $\cardi=1$ & $\ell^*=4$\\
& $S=\{1,2,4\}$ & $\cardi=1$ & $\ell^*=4$\\
\hline
\end{tabular}
\label{table:complete_s_m<=5}
\end{table}

Unfortunately, the converse proofs are based on a case-by-case reasoning, i.e., constructively find a user that can decode a certain number of messages. 
We could not straightforwardly extended these proof to the complete--$S$ PICOD$(\cardi)$ for general $m$.
Here we show proofs of two cases. The other cases can be proved using the similar methods.

\paragraph{Proposition~\ref{prop:achi_opt_m<=5}.Case1}
	We show that for the complete--$S$ PICOD$(1)$ where $S=\{1,3\}$ and $m=5$, the optimal code has length $\ell^*=4$.
%
	We do so by proving the existence of a user with one message in its side information set who can decode the remaining $4$ messages.

	By Proposition~\ref{prop:card=1_complete-s}, there exists a user, say $u_1$, with side information set of size $1$, say $A_1=\{1\}$, who can decode $2$ messages, say $B_1\supseteq\{2, 3\}$.
	User $u_1$ thus can mimic user $u_2$ with side information $A_2=\{1,2,3\}$ and decode its desired message. 
	Therefore user $u_1$ can decode at least 3 messages, $|B_1|\geq 3$.

	Denote the last message that has not been decoded by user $u_1$ as $w_5$. 
	Now, if $w_5$ is desired by some users, i.e., we have a user $u_3$ with $d_3=5$, user $u_1$ can mimic user $u_3$ and decode $w_5$ since $A_3\subset [4]$. Therefore user $u_1$ can decode $4$ messages and $\ell^*\geq 4$.

	Otherwise, $w_5$ is not desired by any users in the system. 
	Since the message that is not desired by any users does not have any effect, by deleting it, the system becomes the complete--$[0:3]$ PICOD$(1)$ with $m=4$. 
	By Theorem~\ref{thm:layer_cont_opt} we have the user with $A=\{5\}$ can decode $4$ messages and
	$\ell^*\geq 4$.

	We apply the achievability for the complete--$\{1,2,3\}$ PICOD$(1)$. 
	This achievability works since $\{1,3\}\subset \{1,2,3\}$.
	By Theorem~\ref{thm:complete_consecutive} we have $\ell^*\leq 4$. 
	This proves the optimality of $\ell^*= 4$ transmissions.

Note: the existence proof based on block cover, as used for Proposition~\ref{prop:power_user_exist}, is also workable for Proposition~\ref{prop:achi_opt_m<=5} as well. 

%

\paragraph{Proposition~\ref{prop:achi_opt_m<=5}.Case2}
	We show that for the complete--$S$ PICOD$(1)$ problem where $S=\{0,2,4\}$ and $m=5$, the optimal code has length $\ell^*=4$.
%
The following lemma, which is a refined version of Proposition~\ref{prop:card=1_complete-s}, is used in the proof.
\begin{lemma}
	\label{lem:exist_power_user_in_group}
	For a complete--$\{s\}$ PICOD$(\cardi)$ with $m$ messages, let $A^\prime\subset [m], |A^\prime|\leq s$, $U_{A^\prime}$ be the group of users who have $A^\prime$ in their side information, i.e., $u_i\in U_{A^\prime}$ if and only if $A^\prime\subseteq A_i$.
	For any $A^\prime$, there exists a user in $U_{A^\prime}$ that can decode at least $\min\{m-s, s+t-|A^\prime|\}$ messages.
	Note: Proposition~\ref{prop:card=1_complete-s} is the case $A^\prime=\emptyset$. 
\end{lemma}
\begin{IEEEproof}[Proof of Lemma~\ref{lem:exist_power_user_in_group}]
	The users in $U_{A^\prime}$ alone can be seen as the users in a new complete--$S^\prime$ PICOD$(\cardi)$, where $S^\prime=\{s-|A^\prime|\}$, $m^\prime=m-|A^\prime|$. 
	By Proposition~\ref{prop:card=1_complete-s} we have that there exists a user in this system that can decode $\min\{s^\prime+\cardi, m^\prime-s^\prime\}=\min\{s+\cardi-|A^\prime|, m-s\}$ messages.
	The above argument holds for all $A^\prime\subset [m], |A^\prime|\leq s$.
\end{IEEEproof}

Back to the proof of Proposition~\ref{prop:achi_opt_m<=5}.Case2.
	We show that by giving one message as a genie, the user with no side information can decode the other $4$ messages.

	Since every user can decode one message, user $u_1$ with $A_1=\emptyset$ can decode message $w_{d_1}$. 
	By Lemma~\ref{lem:exist_power_user_in_group}, we see that there exists a user $u_2\in U_{\{d_1\}}$ that can decode $2$ messages, where $U_{\{d_1\}}$ is the group of users who have side information sets of size $2$ and $w_{d_1}$ in their side information sets.
	Without loss of generality let $A_2=\{d_1, 2\}$ and the two messages that $u_2$ can decode be $w_3, w_4$, $d_1\notin \{2,3,4\}$. 
	Therefore, giving message $w_2$ to user $u_1$ allows it to decode $w_3,w_4$.
	Also, there exists a user with side information $\{d_1,2,3,4\}$ and decodes $w_{d_5}\notin\{d_1,2,3,4\}$. So user $u_1$ can decode $w_{d_5}$ as well.
	Overall, user $u_1$ can decode $4$ messages with the proper genie $w_2$. 
	The code length is therefore lower bounded by $\ell^*\geq 4$.

	For the achievability, we split the users into two groups: $S_1=\{0,2\}$ where users have side information of size $0$ or $2$; $S_2 = \{4\}$ where users have side information of size $4$.
	By Proposition~\ref{prop:max<m/2_nonconsecutive} we can satisfy all users in $S_1$ with $3$ transmission; by Proposition~\ref{prop:card=1_complete-s} we can satisfy all users in $S_2$ with one transmission. 
	In total we use $4$ transmissions to satisfy all users.

\section{Proof of Theorem~\ref{thm:circular-arc_optimality}} 
\label{sec:picod_with_special_network_topology_hypergraph}

In this section, we prove a tight converse bound on the optimal code length for PICOD$(1)$ with \canth. 
We start by introducing some graph theory terminology.

\subsection{Graph Preliminaries} 
\label{sub:graph_prelimiary}

Let $H=(V,\mathcal{E})$ denote a \emph{hypergraph} with vertex set $V$ and edge set $\mathcal{E}$, where an edge $E\in\mathcal{E}$ is a subset of $V$, i.e., $E \subseteq V$.
The hypergraph is called \emph{$r$-uniform} if all edges have cardinality $r$, i.e., $|E|=r, \ \forall E\in \mathcal{E}$. 
For $R\subseteq [|V|]$, the hypergraph is called \emph{$R$-uniform} if all edges have cardinality of some $r\in R$, i.e., $|E|\in R, \ \forall E\in \mathcal{E}$. 
The hypergraph is called \emph{complete $r$-uniform} if all edges with cardinality $r$ exit, i.e., for all $E$ such that $|E|=r, E\subseteq V$, we have $E\in \mathcal{E}$. 
The hypergraph is called complete \emph{$R$-uniform} if all edges with  cardinality $r\in R$ exist. 
The \emph{dual} hypergraph $H^*=(V^*, \mathcal{E}^*)$ of $H$ is a hypergraph where the vertices and edges are interchanged, i.e., $\mathcal{E}^*=V$, $V^*=\mathcal{E}$.

The degree of a vertex $v\in V$ is the number of its incident edges, i.e., $\delta(v)=|\{E: v\in E, E\in\mathcal{E}\}|$. 
The hypergraph is called \emph{$k$-regular} if the degree of all vertices is $k$. 
A \emph{factor} of $H$ is a spanning edge induced subgraph of $H$, i.e., an edge induced subgraph of $H$ with the same vertex set of $V$. A \emph{$k$-factor} is a factor which is $k$-regular.
A hypergraph $H$ is called an \emph{\cah} if there exists an ordering of the vertices $v_1, v_2, \ldots, v_n$ such that if $v_i, v_j, i\leq j$, then the $v_q$ for either all $i\leq q\leq j$, or all $q\leq i$ and $q\geq j$, are incident to an edge $E$.
 
For a PICOD$(\cardi)$, its \nth\ is a hypergraph $H=(V,\mathcal{E})$ such that:
i)   $V=\{u_1,\ldots,u_n\}$, i.e., vertices represent the users;
ii)  $\mathcal{E}=\{E_1,\ldots,E_m\}$, i.e., edges represent the messages;
iii) $u_i\in E_j$ if $w_j\notin A_i$, i.e., a vertex is incident to an edge if the user does not have the message in the side information.
This definition of \nth\ is a generalization of the network topology graph in~\cite{TDMA_optimal}.

Note that the \nth\ is defined solely on user set $U$, message set $W$, and side information sets $\mathcal{A}$.
For the IC, the \nth\ does not uniquely define an instance of the problem, since it does not contain the information about desired message sets of the users.
However, the \nth\ uniquely defines a PICOD$(\cardi)$ due to the property that the PICOD$(\cardi)$ does not specify the desired messages for the users.

\subsection{On the Optimality of a Single Transmission} 
\label{sub:condition_for_one_transmission_to_be_optimal}

We give the necessary and sufficient condition on the \nth\ of a PICOD$(1)$ problem for which one transmission is optimal.
This result applies to all PICOD$(1)$ instances, thus serves as a general converse bound for the PICOD$(1)$.  

\begin{prop}
\label{prop:l=1_condition}
A PICOD$(1)$ with $m$ messages has $\ell^*=1$ if and only if its \nth\ has a $1$-factor. Otherwise $\ell^* \geq 2$.
\end{prop}

\begin{IEEEproof}[Proof of Proposition~\ref{prop:l=1_condition}]

\emph{Achievability}:
The \nth\ $H$ has a $1$-factor if it has an edge induced sub-hypergraph whose vertices are the same as the vertices of $H$ and all have degree one. 
In other words, in this induced sub-hypergraph all vertices are adjacent to one and only one edge.
Since $H$ is the \nth, its vertices represent users and edges represent messages. 
A vertex is adjacent to an edge if and only if the user does not have that message in its side information set. 
For the PICOD$(1)$, that message can be a desired message by the incident users.
Therefore, among all the messages corresponding to the edges in the $1$-factor, every user has one and only one message that is not in its side information set.
Transmitting the sum of all these messages satisfies all users.
By this transmission scheme we achieve $\ell^*= 1$, which is clearly optimal.

\emph{Converse}:
We aim to show that if the \nth\ does not have a $1$-factor hypergraph, then we can construct a user that can decode two messages, thus two transmissions are needed.
For any valid code, consider the sub-hypergraph induced by the edges corresponding to all the desired messages by all users, i.e., the edge induced sub-hypergraph of $H$ where the edges correspond to the messages that are decoded by at least one user.  
This sub-hypergraph is always a factor, i.e., a spanning sub-hypergraph, since all users can decode at least one message in a  PICOD$(1)$. 
Assume no $1$-factor exists in $H$, that is, there exists a vertex whose degree is at least $2$ in the sub-hypergraph.
In other words, for all choices of desired messages at the users, there exists a pair of users $u_1$ and $u_2$ with desired messages $w_{d_1}$ and $w_{d_2}$ such that $d_2\notin A_1$.
We therefore have $A_1 \subseteq [m]\setminus \{d_1, d_2\}$. 
Given any valid code, a user $u^\prime$ with $A^\prime=[m]\setminus \{d_1, d_2\}$ can mimic user $u_1$ then user $u_2$, thus can decode $w_{d_1}, w_{d_2}$. 
By Lemma~\ref{lem:converse_on_decodable_msg}, we conclude that $\ell^*\geq 2$. 
\end{IEEEproof}

\subsection{Proof of Theorem~\ref{thm:circular-arc_optimality}} 
\label{sub:circular_arc_hypergraph}

We show a case where the converse proposed in Proposition~\ref{prop:l=1_condition} is tight by proposing an achievable scheme based on the properties of a \cah. First, in Lemma~\ref{lemma:full_cover} we show the following fact: if two edges, say $E_i$ and $E_j$, are ``close'' in $\mathcal{E}$ with a nonzero gap between them, then there exists an edge in $\mathcal{E}$ that ``covers'' the whole gap between $E_i$ and $E_j$. This fact will be used in Theorem~\ref{thm:circular-arc_optimality} to design a two-transmission achievable scheme. 
\begin{lemma}
\label{lemma:full_cover}
Assume a \cah\ $H$ without isolated vertices and where the vertices are in a cyclic order $\{v_1,v_2,\dots,v_n\}$.
Assume there exist two edges $E_{i}=\{v_{i_1},\ldots,v_{i_p}\}$ and $E_{{j}}=\{v_{j_1},\ldots,v_{j_q}\}$ that satisfy the following two conditions: Condition1) $i_p+1<j_1$, and 
Condition2) every edge that contains any vertices in $\{v_{i_p+1},\dots,v_{j_1-1}\}$ contains $v_{i_p}$. 
Then,  there exists an edge $E_k$ such that $\{v_{i_p+1},\dots,v_{j_1-1}\}\subseteq E_k$.
\end{lemma}
\begin{IEEEproof}[Proof of Lemma~\ref{lemma:full_cover}]
	Since $H$ does not have any isolated vertices, there exists $E_k\in\mathcal{E}$ such that $v_{j_1-1}\in E_k$.
	By the Condition2 we have $v_{i_p}\in E_k$.  
	By the property of \cah\ (if $v_{i_p}$ and $v_{j_1-1}$ are contained in $E_k$, all the vertices between are contained in $E_k$ as well) we have $\{v_{i_p+1},\dots,v_{j_1-1}\}\subseteq E_k$.
\end{IEEEproof}


%

\begin{figure}
  \centering
  \includegraphics[width=0.35\columnwidth]{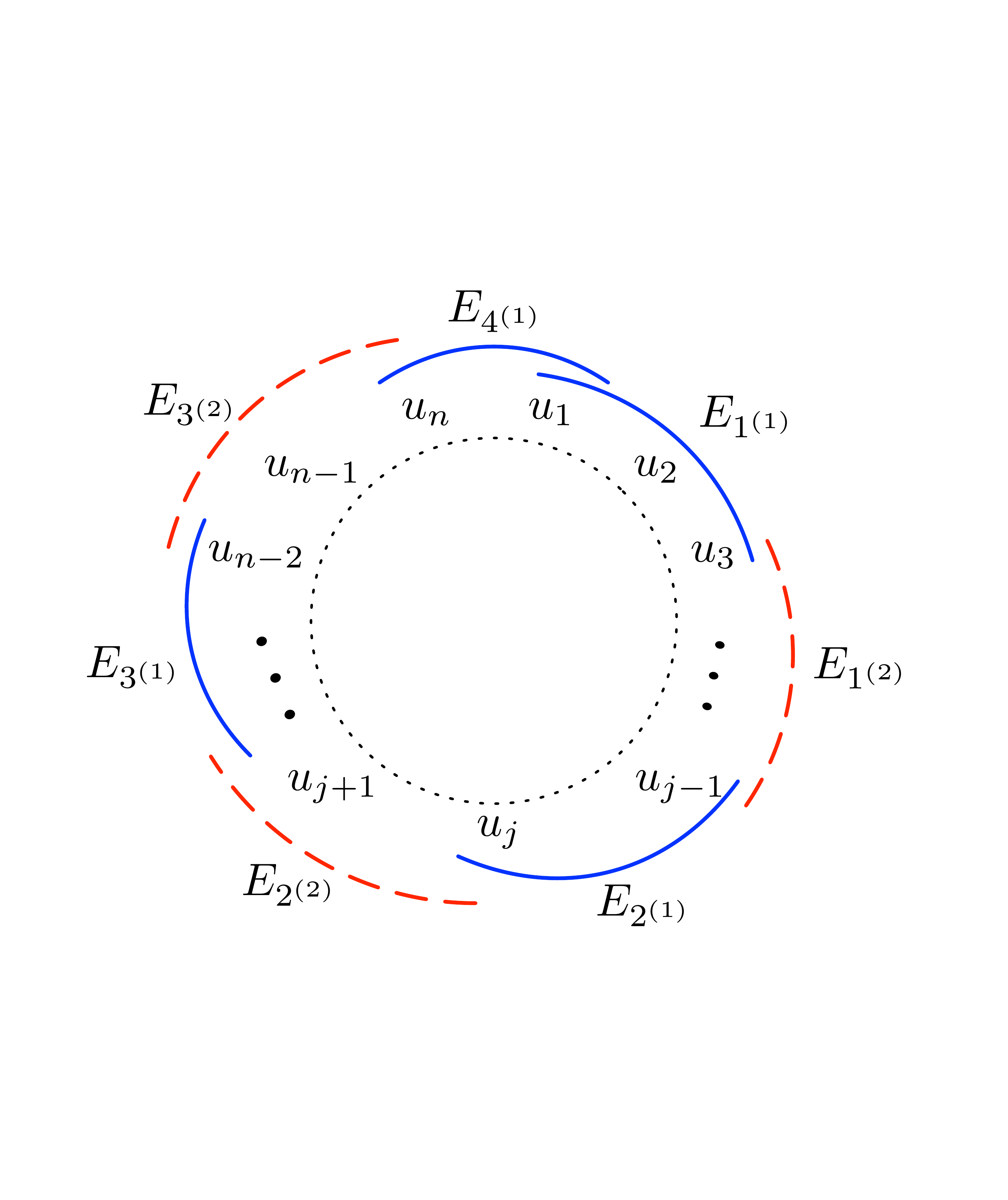}
  \caption{Two transmissions scheme for \canth\ PICOD$(\cardi)$.}
  \label{fig:circular_arc_achievability}
\end{figure}

\begin{IEEEproof}[Proof of Theorem~\ref{thm:circular-arc_optimality}]
%
We propose an achievable scheme that uses two transmissions to satisfy all users for all PICOD$(1)$ instances with \canth. The scheme consists two steps.

\paragraph{Theorem~\ref{thm:circular-arc_optimality}.Step1}
Given a PICOD$(\cardi)$ with \nth as a circular-arc hypergraph, we notice that:
\begin{itemize}
	\item No vertex is isolated.
	\item There might exists an edge that is as a proper subset of another edge.
\end{itemize}
We drop those edges that are proper subsets of the union of other edges, obtaining the edge set $\mathcal{E}$. 
In other words, $|E_i\setminus (\cup_{j\neq i} E_j)|>0, \forall E_i, E_j\in \mathcal{E}$.
The achievability scheme based on $\mathcal{E}$ will be valid for the original problem setting as well (since we are restricted to use less messages to satisfy all users).
The edge induced subgraph by $\mathcal{E}$ has no isolated vertex as well.

\begin{algorithm}
	\SetAlgoLined
	\KwData{User set: $V=\{v_1,\dots, v_n\}$, message set: $\mathcal{E}$.}
	\KwResult{Message set: $\mathcal{E}^{(1)}=\{E_{1^{(1)}}, \dots E_{e^{(1)}}\}$.}
	Initialization: set $i=1$, $\mathcal{E}^{(1)}=\emptyset$.
	\\
	\While{$i\leq n$}{
	Seek an edge that starts at $v_i$, i.e., an edge that is $\{v_i, \dots \}$\;
	\eIf{Such an edge is found}{
	Let $\mathcal{E}^{(1)}$ include be the edge found\;
	$i$ becomes the index of the vertex right after the found edge, that is, the edge $\{\dots, v_{i-1}\}$ \;
	}
	{
	$i=i+1$\;
	}
	}
\caption{Algorithm for finding $\mathcal{E}^{(1)}$ in Step1.}
\label{alg:step1}
\end{algorithm}

In Step1 we find a set of messages $\mathcal{E}^{(1)}\subseteq \mathcal{E}$ by using Algorithm~\ref{alg:step1}.
The blue solid arcs in Fig.~\ref{fig:circular_arc_achievability} show an example of $\mathcal{E}^{(1)}$ found by Algorithm~\ref{alg:step1}.

Denote the cardinality of $\mathcal{E}^{(1)}$ as $e:=|\mathcal{E}^{(1)}|$.
We claim that $\mathcal{E}^{(1)}$ has the following properties:
\begin{itemize}
	\item $E_{i^{(1)}}\cap E_{j^{(1)}} = \emptyset,$ for all $i, j\in [e], (i, j) \neq (1, e)$ and $(i, j) \neq (e, 1)$.
	\item For all $i, j \in [e-1]$, $E_{i^{(1)}}=\{v_{i_1^{(1)}}, \dots , v_{i_{e_i}^{(1)}}\}$, $E_{j^{(1)}}=\{v_{j_1^{(1)}}, \dots , v_{j_{e_j}^{(1)}}\}$, if $i_{e_i}^{(1)}+1 < j_1^{(1)}$, we have an edge $E_{i^{(2)}}\in \mathcal{E}$ such that $\{v_{i_{e_i}^{(1)}}, \dots ,v_{j_1^{(1)}}\}$.
\end{itemize}
The first property holds since the algorithm chooses adjacent edges in $\mathcal{E}^{(1)}$ that are disjoint and there is possibly nonempty intersection between $E_{1^{(1)}}$ and $E_{e}$. The second property holds by Lemma~\ref{lemma:full_cover}.
In the first transmission we send the sum of the messages in $\mathcal{E}^{(1)}$, i.e., $\sum_{i=1}^{e} w_i$. 
The users who are satisfied are in $\cup_{E_i\in \mathcal{E}^{(1)}} E_i \setminus (E_{1^{(1)}} \cap E_{e})$.
In the network topology hypergraph, these are the users that are ``spanned'' by these edges, excluding the users whose vertices are in $E_{1^{(1)}}\cap E_{e}$ where $E_{1^{(1)}}\cap E_{e}\neq \emptyset$. 
Therefore we are left with the users whose	
corresponding vertices are contained in $\left(U\setminus(\cup_{\mathcal{E}^{(1)}} E_{i^{(1)}})\right) \cup \left(E_{1^{(1)}}\cap E_{e}\right)$. 

\paragraph{Theorem~\ref{thm:circular-arc_optimality}.Step2}
The users who are not satisfied by the first transmission are the users whose side information sets contain either all the chosen messages in Theorem~\ref{thm:circular-arc_optimality}.Step1, or both $w_{1^{(1)}}$ and $w_{e}$. 
In other words, in the network topology hypergraph, they are the users who lie ``in between'' the edges, or in the intersection of the first and last edges, in $\mathcal{E}^{(1)}$ chosen in the previous step.

As we have shown in the second property of $\mathcal{E}^{(1)}$ in Theorem~\ref{thm:circular-arc_optimality}.Step1,
for the unsatisfied users between ${E}_{i^{(1)}}\in{\mathcal{E}^{(1)}}$ and ${E}_{ (i+1)^{(1)}}\in{\mathcal{E}^{(1)}}$, there exists an edge ${E}_{i{(2)}}$ that includes all those users.
Therefore, we find a set of edges $\mathcal{E}^{(2)}=\{E_{1^{(2)}},\ldots,E_{{(e-1)}^{(2)}}\}$ such that $U\setminus(\cup_{\mathcal{E}^{(1)}} E_{i^{(1)}})\subseteq \cup_{E_i\in \mathcal{E}^{(2)}}E_i$.
In Fig. \ref{fig:circular_arc_achievability} they are the edges represented by the red dashed arcs.
Note that all edges in $\mathcal{E}^{(2)}$ are pairwise disjoint, since if $E_{i^{(2)}}\cap E_{{i+1}^{(2)}}\neq\emptyset$ then we have $E_{{i}^{(1)}}\subseteq E_{{i-1}^{(2)}}\cup E_{{i}^{(2)}}$, i.e., $|E_{{i}^{(1)}} \setminus (\cup_{j\neq {i}^{(1)}} E_j)|=0$. This is forbidden since we dropped the messages at the beginning of the Step1. 
Moreover, $\left( E_{1^{(1)}}\cap E_{e^{1}} \right) \cap E_{1^{(2)}}=\emptyset$ and $\left( E_{1^{(1)}}\cap E_{e^{1}} \right) \cap E_{(e-1)^{(2)}}=\emptyset$ by the same reasoning.

In the second transmission, we send the sum $\left(\sum_{j=1}^{e-1} w_{j^{(2)}}\right)+w_{1^{(1)}}$.
The users that are not satisfied yet by the first transmission have all but one of the messages in $\{w_{1^{{(2)}}},\dots, w_{{k-1}^{(2)}}, w_{1^{(1)}}\}$ in their side information sets. 
Therefore all the unsatisfied user after Theorem~\ref{thm:circular-arc_optimality}.Step1 can be satisfied by the second transmission.
All the users are satisfied with two transmissions.

This, together with the converse in Proposition~\ref{prop:l=1_condition}, concludes the proof of Theorem~\ref{thm:circular-arc_optimality}.
\end{IEEEproof}

\section{Conclusion and Future Works}
\label{sec:conclusion}

In this paper we provided tight information theoretic converse bounds for some classes of PICOD$(\cardi)$ problems.
The key idea for our converse is to show that for the PICOD$(\cardi)$ with a certain structure of the side information sets, regardless of the choice of desired message sets at the users, there exists a user that can decode a certain number of messages beside its $\cardi$ desired ones.
We showed two methods to prove the existence of such a user: constructive proof and existence proof.
The constructive proof works for the PICOD$(\cardi)$ with \canth, and for the complement-consecutive complete--$S$ PICOD$(\cardi)$ with $m$ messages where $S=[0:m-\cardi]\setminus[\so : \st], 0< \so\leq \st< m-1$.
%
The existence proof works for the consecutive complete--$S$ PICOD$(\cardi)$ with $m$ messages where $S=[\smin : \smax], 0\leq \so\leq \st\leq m-1$.

The key idea for the existence proof was inspired by the similarity of the side information set structure of the consecutive complete--$\{s\}$ PICOD$(\cardi)$ to Steiner systems in combinatorial design. 
Combinatorial design studies the properties of a family of subsets, called blocks, that cover all $s$-element subsets of the same ground set; the results are usually established on the high symmetry of the structure of all $s$-element subsets. 
We introduced the idea of block cover as a tool for the converse proof, together with the classical \text{MAIS} for the IC problem. We solved first the critical complete--$\{s\}$ PICOD$(\cardi)$ with $m=2s+\cardi$ messages, where we showed that a block cover with maximum block size strictly less than $m=2s+\cardi$ does not exist.
%
%
%
For the other considered cases, we showed that we can enhance the system to a critical one.



Open problems and future directions include:
\begin{itemize}

	\item The main contribution of this work are methods to prove the existence of a user that can decode a certain number of messages: constructive and existence proofs.
	While the later shows an advantage over the former on the complexity of the proof, it is based on the strong symmetric structure of the side information set of the users. 
	Like combinatorial design, for the result to hold we need exactly all the $s$-element subsets of ground set $[m]$.
	Therefore, this method suits the complete--$\{s\}$ PICOD$(\cardi)$. 
	For the other cases, we need some extra tools.
	We showed the proof for the consecutive complete--$S$ PICOD$(\cardi)$ by a reduction to the critical case. 
	However, it appears that not all the PICOD$(\cardi)$, even all complete--$S$ PICOD$(\cardi)$, can be reduced in the same fashion without loss of optimality in terms of the code length. 
	Therefore we still lack an efficient method to obtain a general optimal converse bound for the general PICOD$(\cardi)$.
	In Section~\ref{sub:complete_m<=5} we showed the optimality of the proposed achievability up to $m=5$ for the complete--$S$ PICOD$(\cardi)$.  
	The converse is obtained by checking all the cases that are not covered by the Theorem~\ref{thm:layer_cont_opt} or Propositions~\ref{prop:max<m/2_nonconsecutive},~\ref{prop:min>m/2_nonconsecutive},~\ref{prop:min<m/2<max_nonconsecutive}.
	Therefore the method is not systematic and straightforwardly generalizable to general $m$.
	The information theoretical optimal code length for the general complete--$S$ PICOD$(\cardi)$ with $m$ messages is still open.

	\item We notice that in the complete--$S$ PICOD$(\cardi)$ considered in this work, removing/adding some users does not change the optimal code length. 
	In fact, in some cases (e.g., $S=[0:m/2]$) roughly half of the users can be removed without affecting $\ell^*$. 
	These users can be considered as ``non-critical'', 
	in contrast to other ``critical'' users who will change the optimal code length if removed/added. 
	The PICOD$(\cardi)$ is called ``critical'' if all of its users are critical.
	We see the ``critical'' consecutive complete--$S$ PICOD$(\cardi)$ are those with $m\geq \smin+\smax+\cardi$.
	In other words, the ones with ``small'' size of side information/number of desired messages.
	In this case the optimal code length $\smax+\cardi$.
	For this setting, removing any single user reduces the optimal code length by 1. 
	If $m< \smin+\smax+1$, there are $\sum_{s=\smin}^{\smax} \binom{m}{s}- \binom{2m -  2{\smin} -1}{m-{\smin}-1}$ users are non-critical. 
	It is worth to mention that due to the symmetric structure of the complete--$S$ PICOD$(\cardi)$ where $|S|=1$, all users are essentially the same, i.e., all users are critical if any user is critical.
	The question about the critical users in the PICOD$(\cardi)$ is interesting because it shows the redundancy embedded in the system structure. 
	The condition for a complete--$S$ PICOD$(\cardi)$ to be critical, the number of its non-critical users, and in general, the condition to be critical for the general PICOD$(\cardi)$, are the topics of future works for the PICOD$(\cardi)$. 

	\item In the PICOD formulation adopted in this work, the server broadcasts information to all users based on the knowledge all messages in the database. Another practically motivated scenario includes peer-to-peer/distributed models where users broadcast information based on their side information set. The converse bounds developed in this work are also converse bounds for peer-to-peer/distributed model with the same parameters. The open question is whether this ``trivial'' converse bound can be achieved. Surprisingly, it appears that for the consecutive and complement-consecutive complete--$S$ PICOD$(\cardi)$ that we have solved, as long as the problem is ``pliable,'' i.e., there are indeed multiple choices of desired messages that satisfy the users, than the tight results in this paper are tight for the peer-to-peer/distributed model. 
	One of the open questions is to quantify the optimal code length is the non-pliable cases for the complete--$S$ PICOD$(\cardi)$, where the problem reduces to a distributed index coding problem~\cite{distributed_ic}.
	
\end{itemize}

\appendix
\section{Lemma}\label{app:lem:exist_intersection_s}

{
\begin{lemma}
\label{lem:exist_intersection_s}
	For $s+1$ arbitrary subsets $B_i$ 
	from a ground set of size $s$, there exists a set $P\subseteq [s+1]$ such that $|\cap_{i\in P}B_i|=|P|-1$.
\end{lemma}

The proof of Lemma~\ref{lem:exist_intersection_s} is based on induction on $s$ (the size of the ground set in this Lemma) and the following Lemma~\ref{lem:cross_lemma}.

	\begin{lemma}
	\label{lem:cross_lemma}
	    Let $B_1,B_2,\dots,B_x$ are non-empty subsets of set $\{v_1,v_2,\dots,v_y\}$, for some positive integers $x,y$. Let $C_j$ be the collection of subsets that contain $v_j$, i.e., $v_j\in B_i$ if and only if $i\in C_j$. Let $c_j=|C_j|$. 
	    There always exists a pair $(i,j)$ such that $\frac{c_j}{|B_i|}\geq \frac{x}{y}$ and $v_j\in B_i$.
	\end{lemma}

	\begin{IEEEproof}[Proof of Lemma~\ref{lem:cross_lemma}]
	    Construct a $x\times y$ matrix $W$. $w_{ij}=1/|B_i|$ if $v_j\in B_i$, otherwise $w_{ij}=0$. Since $|B_i|\neq 0$ for all $i$, matrix $W$ can be constructed.    
	    Note that the sum of each row is one.
	    We have the summation of all elements in $W$ is $\sum_{i\in[x],j\in[y]}w_{ij}=\sum_{i\in [x]}(\sum_{j\in[y]}w_{ij})=x$, which is the number of rows.
	    The summation of all elements in $W$ can also be obtained by adding up the summation of the columns. Since there are $y$ columns, there exists a column whose summation is no less than the average, i.e., exists $j$ such that  
	    \begin{align}
	        \sum_{k\in[x]}w_{kj} &= \sum_{k:v_j\in B_k}\frac{1}{|B_k|} \geq \frac{x}{y}.
	    \end{align}
	    Let $B_i$ be the smallest subset that contains $v_j$.
	    We have 
	    \begin{align}
	         \sum_{k:v_j\in B_k}\frac{1}{|B_k|} \leq \sum_{k:v_j\in B_k} \frac{1}{|B_i|} = \frac{c_j}{|B_i|}.
	    \end{align}
	    Therefore, for the pair $(i,j)$ we have $v_j \in B_i$ and 
	    \begin{align}
	        \frac{c_j}{|B_i|}\geq \frac{x}{y}.
	    \end{align}
	\end{IEEEproof}

\begin{IEEEproof}[Proof of Lemma~\ref{lem:exist_intersection_s}]
When $|B_i|=0$ for some $i$, take $P=\{i\}$, we have $|\cap_{i\in P}B_i|=0=|P|-1$. Lemma~\ref{lem:exist_intersection_s} is proven. Therefore we just need to consider the case where all $B_i$ are non-empty.

For the initial case $s=1$ the statement in Lemma~\ref{lem:exist_intersection_s} is true. It can be easily seen since $B_1=B_2=\{1\}$ (this is the only $s+1=2$ non empty subsets from a ground set of cardinality $s=1$). Take $P=[2]$; we have $|\cap_{i\in[2]}B_i|=1=2-1$. 

Assume the statement in Lemma~\ref{lem:exist_intersection_s} is true for all $s\leq t-1$. We construct a $P$ such that $|\cap_{i\in P}B_i|=|P|-1$ for $s=t$.
In Lemma~\ref{lem:cross_lemma}, substitute $x$ by $s+1$ and $y$ by $s$, we have a pair $(i,j)$ such that $j\in B_i$ and $\frac{c_j}{|B_i|}\geq \frac{s+1}{s}$, where $c_j=|C_j|$ and $C_j\subseteq [s+1]$ is  the collection of subsets that contain $j$.
By reordering the labels, without loss of generality, let $i=1$ and $B_i=B_1=[j]$. 
Since $\frac{c_j}{|B_1|}\geq \frac{s+1}{s}>1$, we have $c_j>j$, $|C_j\setminus \{1\}|>j-1$.
Consider $B^\prime_{i^\prime}:=B_{i^\prime}\cap[j-1]$, $i^\prime\in C_j\setminus \{1\}$ where $B^\prime_{i^\prime}$ are subsets of $[j-1]$. Since $j-1<s$, by the inductive hypothesis there exists $P^\prime$ such that $|\cap_{i\prime\in P^\prime}B^\prime_{i^\prime}|=|P^\prime|-1$.
Let $P=P^\prime\cup\{1\}$. Note that $j\in B_q$ for all $q\in P$ and $k\notin \cap_{q\in P}B_q$ for all $k\in[j+1:s]$. We have $\cap_{q\in P}B_q=\cap_{j^\prime P^\prime}\cup \{j\}$. Then $|\cap_{q\in P}B_q|=|P^\prime|-1+1=|P|-1$ as $|P|=|P^\prime|+1$.

Therefore we can always find a $P$ such that $|\cap_{i\in P}B_i|=|P|-1$ for all positive integer $s$.
\end{IEEEproof}


\bibliographystyle{IEEEtranS}
\bibliography{refs}

\begin{thebibliography}{10}
\providecommand{\url}[1]{#1}
\csname url@samestyle\endcsname
\providecommand{\newblock}{\relax}
\providecommand{\bibinfo}[2]{#2}
\providecommand{\BIBentrySTDinterwordspacing}{\spaceskip=0pt\relax}
\providecommand{\BIBentryALTinterwordstretchfactor}{4}
\providecommand{\BIBentryALTinterwordspacing}{\spaceskip=\fontdimen2\font plus
\BIBentryALTinterwordstretchfactor\fontdimen3\font minus
  \fontdimen4\font\relax}
\providecommand{\BIBforeignlanguage}[2]{{%
\expandafter\ifx\csname l@#1\endcsname\relax
\typeout{** WARNING: IEEEtranS.bst: No hyphenation pattern has been}%
\typeout{** loaded for the language `#1'. Using the pattern for}%
\typeout{** the default language instead.}%
\else
\language=\csname l@#1\endcsname
\fi
#2}}
\providecommand{\BIBdecl}{\relax}
\BIBdecl

\bibitem{two_wayrelay}
A.~S. Avestimehr, A.~Sezgin, and D.~N. Tse, ``Approximate capacity of the
  two-way relay channel: A deterministic approach,'' \emph{46th Annual Allerton
  Conference on Communication, Control, and Computing}, 2008.

\bibitem{index_coding_with_sideinfo}
Z.~Bar-Yossef, Y.~Birk, T.~S. Jayram, and T.~Kol, ``Index coding with side
  information,'' \emph{IEEE Trans. on Information Theory}, vol.~57, no.~3, pp.
  1479--1494, Mar 2011.

\bibitem{index_coding_original}
Y.~Birk and T.~Kol, ``Informed-source coding-on-demand ({ISCOD}) over broadcast
  channels,'' \emph{Proc. IEEE 17th INFOCOM}, pp. 1257--1264, 1998.

\bibitem{BrahmaFragouli-IT1115-7254174}
S.~Brahma and C.~Fragouli, ``Pliable index coding,'' \emph{IEEE Transactions on
  Information Theory}, vol.~61, no.~11, pp. 6192--6203, Nov 2015.

\bibitem{21np-hard-problems}
R.~M. Karp, ``Reducibility among combinatorial problems,'' \emph{in Complexity
  of Computer Computations}, pp. 85--103, 1972.

\bibitem{constant_frac_satisfactory}
T.~Liu and D.~Tuninetti, ``Pliable index coding: Novel lower bound on the
  fraction of satisfied clients with a single transmission and its
  application,'' \emph{Information Theory Workshop (ITW)}, 2016.

\bibitem{itw_2017}
------, ``Information theoretic converse proofs for some picod problems,''
  \emph{ITW 2017}, 2017.

\bibitem{linear_suboptimality_IC}
E.~Lubetzky and U.~Stav, ``Nonlinear index coding outperforming the linear
  optimum,'' \emph{IEEE Trans. Information Theory}, vol.~55, no.~8, pp.
  3544--3551, August 2009.

\bibitem{IC_NC_matroid}
S.~E. Rouayheb, A.~Sprintson, and C.~Georghiades, ``On the index coding problem
  and its relation to netowrk coding and matroid theory,'' \emph{IEEE Trans.
  Information Theory}, vol.~56, no.~7, pp. 3187--3195, July 2010.

\bibitem{distributed_ic}
P.~Sadeghi, F.~Arbabjolfaei, and Y.-H. Kim, ``Distributed index coding,''
  \emph{Proc. Int. Symp. Inf. Theory}, 2016.

\bibitem{polytime_alg_picod}
L.~Song and C.~Fragouli, ``A polynomial-time algorithm for pliable index
  coding,'' \emph{IEEE Trans. on Information Theory}, vol.~64, no.~2, pp. 979
  -- 999, Feb 2018.

\bibitem{IC_nonshannon}
H.~Sun and S.~A. Jafar, ``Index coding capacity: How far can one go with only
  shannon inequalities?'' \emph{IEEE Trans. on Information Theory}, vol.~61,
  no.~6, pp. 3041--3055, June 2015.

\bibitem{generalized_steiner_system}
J.~H. van Lint, ``On the number of blocks in a generalized steiner system,''
  \emph{Journal of Combinatorial Theory}, vol.~A, no.~80, pp. 353 -- 355, 1997.

\bibitem{TDMA_optimal}
X.~Yi, H.~Sun, S.~A. Jafar, and D.~Gesbert, ``{TDMA} is optimal for all-unicast
  dof region of {TIM} if and only if topology is chordal bipartite,''
  \emph{IEEE Trans. on Information Theory}, vol.~64, no.~3, pp. 2065 -- 2076,
  Mar 2018.

\end{thebibliography}


\end{document}